\documentclass[showpacs,amssymb,10pt,reprint,aps,prd,longbibliography,nofootinbib,floatfix]{revtex4-1}
\usepackage{graphicx,epsfig,amssymb}
\usepackage{amsmath,amsfonts, times}
\usepackage{bm} 

\usepackage{epstopdf}
\usepackage[linktocpage,colorlinks]{hyperref}
\usepackage[caption=false]{subfig}
\usepackage[usenames]{color}     
\usepackage{natbib}
\usepackage{soul}
\usepackage[utf8x]{inputenc}
\usepackage[usenames]{color}

\definecolor{coolblack}{rgb}{0.0, 0.18, 0.39}
\definecolor{darkred}{rgb}{0.5,0,0}
\definecolor{darkgreen}{rgb}{0,0.5,0}
\definecolor{darkblue}{rgb}{0,0,0.5}
\definecolor{lapislazuli}{rgb}{0.15, 0.38, 0.61}
\definecolor{venetianred}{rgb}{0.78, 0.03, 0.08}
\definecolor{bleudefrance}{rgb}{0.19, 0.55, 0.91}
\definecolor{dogwoodrose}{rgb}{0.84, 0.09, 0.41}
\hypersetup{colorlinks=true, citecolor=darkgreen, linkcolor=darkblue, 
	urlcolor = blue}

\newcommand\numberthis{\addtocounter{equation}{1}\tag{\theequation}}

\def\be{\begin{equation}}
\def\ee{\end{equation}}

\newcommand{\bea}{\begin{eqnarray}}
\newcommand{\eea}{\end{eqnarray}}
\newcommand{\ben}{\begin{enumerate}}
	\newcommand{\een}{\end{enumerate}}
\newcommand{\bi}{\begin{itemize}}
	\newcommand{\ei}{\end{itemize}}

\def\ga{\mathrel{\raise.3ex\hbox{$>$\kern-.75em\lower1ex\hbox{$\sim$}}}}
\def\la{\mathrel{\raise.3ex\hbox{$<$\kern-.75em\lower1ex\hbox{$\sim$}}}}

\def\be{\begin{equation}}
\def\ee{\end{equation}}

\def\I_M{{I_{\scriptscriptstyle M\times M}}}

\def\be{\begin{equation}}
\def\ee{\end{equation}}
\def\bea{\begin{eqnarray}}
\def\eea{\end{eqnarray}}
\newcommand{\beq}{\begin{eqnarray}}
\newcommand{\eeq}{\end{eqnarray}}

\begin{document}
	\title{\large Compact objects in quadratic Palatini gravity generated by a free scalar field}

	\author{Renan B. Magalh\~aes}
	\email{renan.magalhaes@icen.ufpa.br}
	\affiliation{Programa de P\'os-Gradua\c{c}\~{a}o em F\'{\i}sica, Universidade 
		Federal do Par\'a, 66075-110, Bel\'em, Par\'a, Brazil.}
	
	\author{Lu\'is C. B. Crispino}
	\email{crispino@ufpa.br}
	\affiliation{Programa de P\'os-Gradua\c{c}\~{a}o em F\'{\i}sica, Universidade 
		Federal do Par\'a, 66075-110, Bel\'em, Par\'a, Brazil.}
	
	\author{Gonzalo J. Olmo}
	\email{gonzalo.olmo@uv.es}
	\affiliation{Departamento de F{\'i}sica Te{\'o}rica and \textit{IFIC}, Centro Mixto Universidad de Valencia - \textit{CSIC}. Universidad de Valencia, Burjassot-46100, Valencia, Spain.}

	\begin{abstract}
		We study the correspondence that connects the space of solutions of General Relativity (GR) with that of Ricci-based Gravity theories (RBGs) of the $f(R,Q)$ type in the metric-affine formulation, where $Q=R_{(\mu\nu)}R^{(\mu\nu)}$. We focus on the case of scalar matter and show that when one considers a free massless scalar in the GR frame, important simplifications arise that allow to establish the correspondence for arbitrary $f(R,Q)$  Lagrangian. We particularize the analysis to a quadratic $f(R,Q)$ theory and use the spherically symmetric, static solution of Jannis-Newman-Winicour as seed to generate new compact objects in our target theory. We find that two different types of solutions emerge, one representing naked singularities and another corresponding to asymmetric wormholes with bounded curvature scalars everywhere. The latter solutions, nonetheless, are geodesically incomplete. 
	\end{abstract}
	
	\date{\today}
	
	\maketitle

\section{Introduction}\label{sec:int}

The recent observational achievements in strong gravity scenarios provided by gravitational waves detections~\cite{ligo,Abbott2016GW151226} and the imaging of a hot plasma accretion disk around the central object of the galaxy M87 \cite{eht} are stimulating new ideas and research lines in the gravitation and astrophysics communities. The collection of gravitational waves measured so  far~\cite{Abbott2017GW170104,Abbott2017GW170814,Abbott2017GW170608}  brings in new information that allows to explore in some detail key aspects of compact objects that go beyond the paradigm set by black holes (BHs)~\cite{
Herdeiro:2014,Herdeiro:2016,Delgado:2021}. 
Moreover, the view provided by advanced numerical simulations and the very large baseline interferometry techniques that enabled the reconstruction of the shadow of the supermassive object in the center of M87  \cite{Gold:2020iql} stands now as a solid complementary source of information for the study of critical light curves and light rings \cite{Narayan:2019imo,Luminet:1979nyg,Falcke:1999pj,Cunha:2016,Cunha:2017,Cunha:2020,LimaJunior:2021}, which are crucial elements to tell apart BHs from other exotic compact objects.

Already within general relativity (GR), one can construct various kinds of compact objects~\cite{Mielke,Mazur,visser,Damour,Cardoso:2016,Junior:2020,Herdeiro:2021JCAP} that can mimic some BH features, turning the problem of distinguishing BHs from those mimickers at infinity~\cite{Cardoso:2019rvt,lemos} into a technically challenging task. Moreover, alternative theories of gravity may generate even more exotic types of compact objects with some astrophysical properties which are close or even coincident with those of BHs~\cite{KZ}. In general, however, these alternative solutions may still exhibit distinguishing features in the (expected) near horizon region, thus providing challenging opportunities to tell them apart from standard BHs via gravitational waves \cite{Ezquiaga:2020dao} and/or shadows \cite{Wielgus:2021peu}.
 Hence, it turns out to be very important to better understand these classes of alternative compact objects and their defining properties in order to define strategies that may help to extract key informations from current and future strong gravity observational data.

Standard matter sources that verify the energy conditions can generate BH mimickers in GR and beyond. However, the extra freedom provided by alternative gravity theories has a price to pay, which typically involves more complicated field equations and/or new degrees of freedom, requiring considerable extra efforts to find solutions. Thus, solving the corresponding field equations usually implies cumbersome calculations that make it difficult, and sometimes impossible, to find analytical expressions for the fields associated to the alternative theories of gravity coupled to matter sources. In this sense, a correspondence between the solution spaces of GR and a certain class of alternative gravity theories has been recently established~\cite{AORG2018}, opening a new avenue to explore both analytical and numerical solutions of modified gravity theories from within GR itself. This correspondence, which applies to metric-affine theories in which the gravity Lagrangian is some function of the symmetric part of the Ricci tensor (Ricci-Based Gravity theories, or RBGs for short) allows to establish a map between solutions of GR coupled to some matter Lagrangian and solutions of an alternative theory of gravity coupled to a (usually non-linear) matter Lagrangian~\cite{Afonso:2019fzv}. Explicit examples of this correspondence and applications to scenarios with scalar fields, electromagnetic fields, and fluids have been worked out for $f(R)$ theories and also for the Born-Infeld gravity model (EiBI) in Refs.~\cite{Afonso:2018hyj,Afonso:2018mxn,Afonso:2019fzv,Delhom:2019zrb,Afonso:2021pga}. It should be noted that this correspondence somehow extends beyond the vacuum case the deep relation already observed in \cite{Ferraris:1992dx,Borowiec:1996kg}  between GR and these metric-affine theories.

In this paper we implement the mapping method just mentioned in another relevant family of gravity theories of the $f(R,Q)$ form formulated \textit{a la} Palatini, where $R$ is the contraction of the Ricci tensor with the metric and $Q\equiv  R_{(\mu\nu)}R^{(\mu\nu)}$. We particularize our analysis to the quadratic gravity case, characterized by $R^2$ and $R_{(\mu\nu)}R^{(\mu\nu)}$ corrections to GR~\cite{Olmo:2009,Barragan:2010,Olmo:2012nx}. This type of theory captures essential aspects of semiclassical approaches to quantum gravity, where a Riemann squared term is usually replaced by $R^2$ and $R_{\mu\nu}R^{\mu\nu}$ contributions.  Moreover, focusing on the RBG part of quadratic gravity, one guarantees the projective invariance of the theory, which is crucial to avoid ghost-like instabilities \cite{BeltranJimenez:2019,BeltranJimenez:2020sqf}. This metric-affine version of quadratic gravity (as well as all $f(R,Q)$ metric-affine theories), therefore, is ghost free, unlike its purely metric counterpart. We will focus on the coupling of this gravity theory to scalar matter, paralleling the procedure introduced in Ref.~\cite{AORG2018} to investigate static, spherically symmetric objects, where solutions for $f(R)$ and the EiBI model were obtained. As we will see,  this quadratic model yields solutions that are in qualitative agreement with those found in Ref.~\cite{AORG2018} but also with those obtained in Ref.~\cite{Afonso:2017aci}, using a different approach. This qualitative agreement confirms the existence of generic properties of scalar compact objects in this type of metric-affine theories of gravity, which suggests that an effective phenomenological description of such objects might be possible. 

The remaining of this paper is organized as follows. In Sec.~\ref{sec:QPG} we introduce the RBG model which we will use, and derive its field equations. The mapping procedure between GR and $f(R,Q)$ gravity using a free scalar field is presented in Sec.~\ref{c_map}. In Sec.~\ref{sec:fRQQuad} we apply the mapping algorithm to the quadratic $f(R,Q)$, and present the scalar objects generated by this procedure in Sec.~\ref{sec:sca_conf}. Finally, we summarize our results and discuss some perspectives in Sec.~\ref{sec:summ}.

\section{Model and field equations} \label{sec:QPG}
The study of $f(R)$ and other extensions of GR are nowadays motivated mainly by phenomenological reasons inspired by the need to unveil new physics at high- and low-energy scales in relation with compact objects, the early universe, and the cosmic speedup problem (see Refs.~\cite{Saridakis:2021vue,CANTATA:2021ktz,Harko:2018ayt,olmo:2011paf,Starobinsky:1980te} and references therein). Another relevant example of modified gravity theory within the metric-affine framework is represented by the Eddington-inspired Born-Infeld model~\cite{BeltranJimenez:2017doy,banados2010,Vollick:2003qp}, which emulates the well-known Born-Infeld electrodynamics theory in an attempt to improve the gravitational dynamics at high energies, by setting bounds on the curvature invariants via a determinantal type action. A power series expansion of this model is dominated by quadratic corrections with specific coefficients in front of the $R^2$ and $R_{(\mu\nu)}R^{(\mu\nu)}$ terms. Exploring quadratic extensions of GR with arbitrary coefficients is thus relevant to determine the robustness of the predictions of such models and also as a way to better understand the phenomenology of models inspired by the semiclassical approach to quantum gravity, as the quadratic terms are the leading order corrections in a power series expansion. It is for this reason that here we consider a theory described by the action
\begin{equation}
S[g,\Gamma,\psi_{m}]=\dfrac{1}{2\kappa^2}\int\text{d}^4x\sqrt{-g}f(R,Q) + S_{m}[g,\psi_{m}],
\label{eq:action}
\end{equation}
where the gravity Lagrangian $\mathcal{L}_{G}=f(R,Q)/(2\kappa^2)$ is a function of $R$ and $Q=g^{\mu\alpha}g^{\nu\beta}R_{(\mu\nu)}R_{(\alpha\beta)}$, and $S_m$ is the matter action, given by
\begin{equation}
\label{eq:lag_matter}
S_m[g_{\mu\nu},\psi_{m}]=\int\text{d}^4x\,\sqrt{-g}\mathcal{L}_{m}(g_{\mu\nu},\psi_{m}),
\end{equation}
where $\mathcal{L}_{m}$ is the matter Lagrangian density, which is a function of the matter fields $\psi_{m}$ and the spacetime metric $g_{\mu\nu}$.

To obtain the field equations of this theory, we will assume a metric-affine (or Palatini) framework, in which the metric field $g_{\mu\nu}$ and the connection $\Gamma^{\alpha}_{\,\,\mu\nu}$ are treated as two independent gravitational fields, so that $R_{(\mu\nu)}$ (for now on we will remove the parenthesis of the  symmetrized Ricci tensor) is a function only of the connection. As is now well established and we will see below, in nonlinear Palatini extensions of GR the connection is not the usual Levi-Civita connection of the metric, as it has to satisfy a specific set of model-dependent equations. As such, the non-metricity tensor  $\nabla^{\Gamma}_{\,\,\mu}\,g_{\alpha\beta}$ is typically non-vanishing. {Torsion, on the other hand, can be trivialized if fermions are coupled via de Dirac equation, i.e., if there is no explicit coupling between them and the axial part of torsion, as it simply becomes a gauge degree of freedom that can be eliminated by a suitable choice of projective gauge. If one chooses to couple fermions and torsion, then the matter sector generates a contribution to the hypermomentum upon variation with respect to the connection, which must be considered explicitly for the consistent resolution of the connection equations. For bosonic fields, torsion can also be gauged away when they are minimally coupled to the metric, while for nonminimal couplings one generally expects a non-trivializable contribution to the hypermomentum \cite{Afonso:2017bxr}.  Here we will focus on minimally coupled scalar fields. }

\subsection{Field equations}
To obtain the field equations, we start by performing a usual variation of the action~\eqref{eq:action}, namely, 
\begin{align*}
\delta S =\dfrac{1}{2\kappa^2}&\int\text{d}^4x\Big[-\dfrac{1}{2}\sqrt{-g}fg_{\mu\nu}\delta g^{\mu\nu} \\&+\sqrt{-g}f_{R}(R_{\mu\nu}\delta g^{\mu\nu} + g^{\mu\nu}\delta R_{\mu\nu}) \\&+ 2\sqrt{-g}f_{Q}(R_{\mu\alpha}R^{\alpha}_{\,\,\nu}\delta g^{\mu\nu}+R^{\mu\nu}\delta R_{\mu\nu})\Big] + \delta S_{m}, \numberthis
\label{eq:variation_action_pala1}
\end{align*}
where $f_{R} = \tfrac{\partial f}{\partial R}$ and $f_{Q} = \tfrac{\partial f}{\partial Q}$. 
We can rearrange the integral in Eq.~\eqref{eq:variation_action_pala1} as
\begin{equation}
\label{eq:variation_action_integral}
I = \delta I_{g} + \delta I_{\Gamma},
\end{equation}
where
\begin{align}
\label{eq:Ig} \delta I_{g} = &\dfrac{1}{2\kappa^2}\int\text{d}^4x\sqrt{-g}\Big(f_{R}R_{\mu\nu} - \dfrac{f}{2}g_{\mu\nu} \\&+ 2f_{Q}R_{\mu\alpha}R^{\alpha}_{\,\,\nu}\Big)\delta g^{\mu\nu}, \\
\label{eq:Igamma} \delta I_{\Gamma} = &\dfrac{1}{2\kappa^2}\int\text{d}^4x\sqrt{-g}\left(f_{R}g^{\mu\nu}+2f_{Q}R^{\mu\nu}\right)\delta R_{\mu\nu}.
\end{align}
We recall that $\delta R_{\mu\nu}$ is given by 
\begin{equation}
\label{eq:var_Ricci_connection2}
\delta R_{\mu\nu} = \nabla_{\rho}\delta\Gamma^{\rho}_{\nu\mu} - \nabla_{\nu}\delta\Gamma^{\rho}_{\rho\mu} + 2S^{\lambda}_{\rho\nu}\delta\Gamma^{\rho}_{\lambda\mu},
\end{equation}
where $2S^{\lambda}_{\mu\nu}\equiv\Gamma^{\lambda}_{\mu\nu} - \Gamma^{\lambda}_{\nu\mu}$ is the torsion tensor. 
Neglecting all torsional terms for simplicity (see Ref.~\cite{Afonso:2017bxr} for details), Eq.~\eqref{eq:Igamma} reduces to
\begin{align*}
\label{eq:Igamma1} 
\delta I_{\Gamma} =\dfrac{1}{2\kappa^2}\int &\text{d}^4x\sqrt{-g}\left(f_{R}g^{\mu\nu}+2f_{Q}R^{\mu\nu}\right)\\&\times \left( \nabla^{\Gamma}_{\rho}\delta\Gamma^{\rho}_{\nu\mu} - \nabla^{\Gamma}_{\nu}\delta\Gamma^{\rho}_{\rho\mu}\right) \numberthis
\end{align*}
and integrating by parts we get
\begin{align*}
\delta I_{\Gamma} =&\dfrac{1}{2\kappa^2}\int\text{d}^4x\nabla^{\Gamma}_{\,\,\rho}\left[\sqrt{-g} J^{\rho}\right]\\&-\dfrac{1}{2\kappa^2}\int\text{d}^4x\delta\Gamma^{\rho}_{\nu\mu}\Big\{\nabla^{\Gamma}_{\rho}\left[\sqrt{-g}\left(f_{R}g^{\mu\nu}+2f_{Q}R^{\mu\nu}\right)\right]\\&-\nabla^{\Gamma}_{\theta}\left[\sqrt{-g}\left(f_{R}g^{\mu\theta}+2f_{Q}R^{\mu\theta}\right)\delta^{\nu}_{\rho}\right]\Big\}, \numberthis
\label{eq:Igamma2} 
\end{align*}
where $J^{\rho}$ is
\begin{equation}
\label{eq:Jrho}
J^{\rho} = \left(f_{R}g^{\mu\nu}+2f_{Q}R^{\mu\nu}\right)\left(\delta \Gamma^{\rho}_{\nu\mu} - \delta^{\rho}_{\nu}\delta \Gamma^{\sigma}_{\sigma\mu}\right).
\end{equation}
Using that $\nabla^{\Gamma}_{\,\,\rho}\sqrt{-g} = \partial_{\rho}\sqrt{-g} - \Gamma^{\sigma}_{\,\,\rho\sigma}\sqrt{-g}$, the first integral of Eq.~\eqref{eq:Igamma2} can be written as 
\begin{equation}
\dfrac{1}{2\kappa^2}\int\text{d}^4x\nabla^{\Gamma}_{\,\,\rho}\left[\sqrt{-g} J^{\rho}\right] = \dfrac{1}{2\kappa^2}\int\text{d}^4x\,\partial_{\rho}\left(\sqrt{-g}J^{\rho}\right) \ .
\label{eq-der}
\end{equation}
Neglecting boundary terms, Eq.~\eqref{eq:Igamma2} becomes
\begin{align*}
\label{eq:Igamma3} 
\delta I_{\Gamma}=&-\dfrac{1}{2\kappa^2}\int\text{d}^4x\delta\Gamma^{\rho}_{\nu\mu}\Big\{\nabla^{\Gamma}_{\rho}\left[\sqrt{-g}\left(f_{R}g^{\mu\nu}+2f_{Q}R^{\mu\nu}\right)\right]\\&-\nabla^{\Gamma}_{\theta}\left[\sqrt{-g}\left(f_{R}g^{\mu\theta}+2f_{Q}R^{\mu\theta}\right)\delta^{\nu}_{\rho}\right]\Big\},\numberthis
\end{align*}
{which is identically zero when one takes the trace over $\rho$ and $\nu$, but for $\rho\neq\nu$ yields}
\begin{equation}
\label{eq:Igamma4} 
\delta I_{\Gamma}=-\dfrac{1}{2\kappa^2}\int\text{d}^4x\delta\Gamma^{\rho}_{\nu\mu}\nabla^{\Gamma}_{\rho}\left[\sqrt{-g}\left(f_{R}g^{\mu\nu}+2f_{Q}R^{\mu\nu}\right)\right].
\end{equation} 
Hence, by using Eqs.~\eqref{eq:Ig} and~\eqref{eq:Igamma4}, the variations with respect to the metric and the connection can be written as
\begin{align}
\label{eq:variation_action_metric}f_{R}R_{\mu\nu} - \dfrac{f}{2}g_{\mu\nu} + 2f_{Q}R_{\mu\alpha}R^{\alpha}_{\,\,\nu} &= \kappa^2T_{\mu\nu},\\
\label{eq:variation_action_connection}\nabla^{\Gamma}_{\,\,\rho}\Big[\sqrt{-g}(f_{R}g^{\mu\nu}+2f_{Q}R^{\mu\nu})\Big]&=0,
\end{align}
respectively, where the stress-energy tensor $T_{\mu\nu}$ is given by 
\begin{equation}
\label{eq:def_Tmn}
T_{\mu\nu}=\dfrac{-2}{\sqrt{-g}}\dfrac{\delta S_{m}}{\delta g^{\mu\nu}}.
\end{equation}
The connection equation \eqref{eq:variation_action_connection} can be formally expressed as 
\begin{equation}
\nabla^{\Gamma}_{\,\,\rho}\Big[\sqrt{-h}h^{\mu\nu}\Big]=0 \ , 
\label{eq:levi_civita}
\end{equation}
where we introduced an {\it auxiliary metric} $h_{\mu\nu}$ which is related to the spacetime metric $g_{\mu\nu}$ by
\begin{align}
\label{eq:metric_relation1}h_{\mu\nu} &= g_{\mu\alpha}{\Omega^\alpha}_\nu \\
\label{eq:metric_relation2} h^{\mu\nu} &= {{[\Omega^{-1}]}^\mu}_\alpha g^{\alpha \nu} \ ,
\end{align}
where 
\begin{eqnarray}
{\Omega^{\alpha}}_\nu&=&\sqrt{|\hat{\Sigma}|}\,\big(\Sigma^{-1}\big)^{\alpha}_{\,\,\nu}  \label{eq:Omega_sigma} \, ,\\ 
{\Sigma^\alpha}_\nu &=& f_{R}{\delta^\alpha}_\nu + 2f_{Q}g^{\alpha\sigma}R_{\sigma\nu} \, ,
\label{eq:sigma_tensor}
\end{eqnarray} 
and $|\hat{\Sigma}|$ denotes the determinant of ${\Sigma^\alpha}_\nu$, as defined in Eq.~\eqref{eq:sigma_tensor} (note that  $|\hat{\Sigma}|=|\hat{\Omega}|$). With this notation, the solution to Eq.~\eqref{eq:levi_civita} can be formally written as the Levi-Civita connection of the metric $h_{\mu\nu}$, namely
\begin{equation}
\label{eq:christoffel_symbol_metric_h}
\Gamma^{\alpha}_{\mu\nu} = \dfrac{1}{2}h^{\alpha\beta}\left(\partial_{\mu}h_{\nu\beta}+\partial_{\nu}h_{\mu\beta}-\partial_{\beta}h_{\mu\nu}\right).
\end{equation}
In order to find the relation that links $h_{\mu\nu}$ with $g_{\mu\nu}$, it is now useful to raise the first index of Eq.~\eqref{eq:variation_action_metric} with $g^{\mu\alpha}$ and introduce the object\footnote{Note that ${P^\mu}_\nu$ is a {\it hybrid} object that is constructed with the inverse metric $g^{\mu\alpha}$ and the Ricci tensor of the independent connection $\Gamma$. This is why we denote it with a new letter.} ${P^\mu}_\nu\equiv g^{\mu\alpha}R_{\alpha\nu}(\Gamma)$ to rewrite Eq.~\eqref{eq:variation_action_metric}  as
\begin{equation}
f_{R}{P^\mu}_\nu -\dfrac{f}{2}{\delta^\mu}_\nu+ 2f_{Q}{P^\mu}_\alpha{P^\alpha}_\nu = \kappa^2{T^\mu}_\nu \ , \\
\label{eq:P_tensor_eq}
\end{equation}
which, in matrix form, reads (here a hat denotes matrix)
\begin{equation}
f_{R}\hat{P} -\dfrac{f}{2}\hat{I}+ 2f_{Q}\hat{P}^2 = \kappa^2\hat{T} \, .
\label{eq:P_matrix_eq}
\end{equation}
 Elementary manipulations allow to express this equation as 
\begin{equation}
\Big(\hat{P}+\dfrac{1}{4}\dfrac{f_{R}}{f_{Q}}\hat{I}\Big)^2 =\left[\left(\dfrac{f_{R}}{4f_{Q}}\right)^2+\dfrac{f}{4f_{Q}}\right]\hat{I} + \frac{\kappa^2}{2f_Q}\hat{T}.
\label{eq:square_completed}
\end{equation}
It is very important to note that Eqs.~\eqref{eq:P_matrix_eq} and~\eqref{eq:square_completed} imply that one can find an expression for the matrix $\hat{P}$ in terms of the scalars $R$ and $Q$ and of the matrices $\hat I$ and $\hat T$, namely, $\hat{P}=\hat{P}(R,Q;\hat I,\hat T)$. Taking the trace of this formal relation, one gets an expression for the scalar $R=P^{\mu}_{\,\,\mu}$, while 
taking the trace of its square one gets an expression for $Q=P^{\mu}_{\,\,\alpha}P^{\alpha}_{\,\,\mu}$. By combining those two relations one can, in principle, obtain expressions for $R$ and $Q$ in terms of the matter variables only. Thus, in general, for any $f(R,Q)$ theory one can conclude that the scalars $R$ and $Q$ can always be written as functions of the matter sources, i.e., $R=R(\hat{T})$ and $Q=Q(\hat{T})$. This conclusion also extends to non-scalar quantities such as $\hat P$ itself and the deformation matrix $\hat{\Omega}$, defined in Eq.~(\ref{eq:Omega_sigma}), which must be regarded as functions of $\hat T$. 

An almost trivial consequence of the above is that, in vacuum, we have that $\hat{T}=0$ and Eq.~\eqref{eq:P_matrix_eq} reduces to
\begin{equation}
\Big(\hat{P}+\dfrac{1}{4}\dfrac{f_{R}}{f_{Q}}\hat{I}\Big)^2 =\left[\left(\dfrac{f_{R}}{4f_{Q}}\right)^2+\dfrac{f}{4f_{Q}}\right]\hat{I} ,
\label{eq:P_matrix_eq_vac}
\end{equation}
{which admits solutions of the form (see Ref.~\cite{BeltranJimenez:2020guo} for a discussion of other possible solutions in this type of equations)}
\begin{equation}
\hat{P}=\Lambda \hat{I},
\label{eq:P_matrix_I}
\end{equation}
where $\Lambda\equiv \Lambda(R^{\text{vac}},Q^{\text{vac}})$ is a constant determined by the vacuum spacetime scalars, $R^{\text{vac}}$ and $Q^{\text{vac}}$. Indeed, Eq.~\eqref{eq:P_matrix_I} states that $R^{\text{vac}} = \Lambda/4$ and $Q^{\text{vac}} = \Lambda^2/4$ are constants, so that the deformation matrix $\hat{\Omega}$ is also constant and proportional to the identity matrix. Hence, the vacuum field equations~\eqref{eq:P_matrix_eq_vac} are the same as those of GR in vacuum with a cosmological constant $\Lambda$, which depend on the gravity Lagrangian $f=f(R,Q)$ chosen. Clearly, this implies that the only propagating degrees of freedom in vacuum are the standard two polarizations of GR, which makes these theories compatible with the LIGO-Virgo constraints on the speed of gravitational waves \cite{ligo}.\\

Before concluding this section, let us note that Eq.~\eqref{eq:P_tensor_eq} can also be written as 
\begin{equation}
f_{R}{P^\mu}_\nu + 2f_{Q}{P^\mu}_\alpha{P^\alpha}_\nu = \kappa^2{T^\mu}_\nu +\dfrac{f}{2}{\delta^\mu}_\nu\ , \\
\label{eq:P_tensor_eq2}
\end{equation}
such that the left-hand side is equivalent to 
\begin{equation}
{\Sigma^\mu}_\alpha{P^\alpha}_\nu=|\Omega|^{1/2} {{[\Omega^{-1}]}^\mu}_\alpha g^{\alpha \beta}R_{\beta\nu}=|\Omega|^{1/2} h^{\mu\beta}R_{\beta\nu}(\Gamma) \ .
\end{equation}
 Given that $R_{\beta\nu}(\Gamma)$ is defined in terms of $\Gamma$ [see Eq.~\eqref{eq:christoffel_symbol_metric_h}], it is evident that $R_{\beta\nu}(\Gamma)=R_{\beta\nu}(h)$, i.e., it is the Ricci tensor of the metric $h_{\mu\nu}$, and $h^{\mu\beta}R_{\beta\nu}(\Gamma)=h^{\mu\beta}R_{\beta\nu}(h)={R^\mu}_\nu(h)$. With this, we can finally rewrite Eq.~\eqref{eq:P_tensor_eq2} as 
\begin{equation}
{R^\mu}_\nu(h)= \dfrac{\kappa^2}{\sqrt{|\hat{\Omega}|}}\Big(\mathcal{L}_{G}\delta^{\mu}_{\,\,\nu}+T^{\mu}_{\,\,\nu}\Big) \ ,
\label{eq:ricci_h}
\end{equation}
where $\mathcal{L}_{G}=f(R,Q)/2\kappa^2$ is the gravity Lagrangian. Using this result, it is straightforward to derive an equation for the Einstein tensor ${G^\mu}_\nu(h)\equiv{R^\mu}_\nu(h) - {\delta^\mu}_\nu(h)R^\alpha_\alpha(h)/2$ of $h_{\mu\nu}$ as 
\begin{equation}
{G^\mu}_\nu(h)= \dfrac{\kappa^2}{\sqrt{|\hat{\Omega}|}}\Big({T^\mu}_\nu-\dfrac{1}{2}{\delta^\mu}_\nu\big(T+2\mathcal{L}_{G}\big)\Big) \ .
\label{eq:Einstein_frame}
\end{equation}
Equation~\eqref{eq:Einstein_frame} is a second-order differential equation for the auxiliary metric $h_{\mu\nu}$, with its left-hand side representing the Einstein tensor of the metric $h_{\mu\nu}$ and the right-hand side describing the matter sources (which are coupled to $g_{\mu\nu}$). This representation of the field equations is not unique to $f(R,Q)$ theories but a general feature of all Ricci-Based Gravity theories (RBGs).  Recall that since $R$ and $Q$ are functions of ${T^\mu}_\nu$, it follows that $|\hat{\Omega}|$ and $\mathcal{L}_G$ are also determined by the matter distribution. 

\section{Map between gravity theories: From GR to $f(R,Q)$ theories}
\label{c_map}
In order to find solutions to Eq.~\eqref{eq:Einstein_frame}, the direct integration approach is not, in general, a good option because its right-hand side is written in terms of the matter sources coupled to the metric $g_{\mu\nu}$, while the left-hand side is specified by $h_{\mu\nu}$, which complicates the analysis. However, a useful strategy has been recently introduced \cite{AORG2018} to deal with the field equations of RBGs, as if one were dealing directly with GR. The key point consists in writing the right-hand side of Eq.~\eqref{eq:Einstein_frame} in terms of a modified matter Lagrangian minimally coupled to the metric $h_{\mu\nu}$, removing in that way any reference to the spacetime metric $g_{\mu\nu}$. This method has been used to find various types of solutions in  $f(R)$ theories and in the Eddington-inspired Born-Infeld gravity theory \cite{Afonso:2019fzv,Afonso:2018hyj,Afonso:2018mxn,Delhom:2019zrb,Afonso:2021pga}. In this section, we will focus on implementing the map between GR and $f(R,Q)$  gravity theories in order to facilitate the construction of new solutions in a larger and different family of theories. For concreteness, we will consider a real scalar field as matter source and at some point we will focus on the free massless case, as key simplifications arise in the equations involved. \\

\subsection{Preliminaries: $\hat\Omega$ in different frame variables}
Following the analysis presented in Ref.~\cite{AORG2018}, we start by rewriting the right-hand side of Eq.~\eqref{eq:Einstein_frame} as a standard stress-energy tensor,
\begin{equation}
{G^\mu}_\nu(h)= \kappa^2{\tilde{T}^\mu}_{\ \ \nu}(h) \ ,
\label{eq:Einstein_frame_GR}
\end{equation}
such that 
\begin{equation}
{\tilde{T}^\mu}_{\ \ \nu}(h)=  \dfrac{1}{\sqrt{|\hat{\Omega}|}}\Big({T^\mu}_\nu-\dfrac{1}{2}{\delta^\mu}_\nu\big(T+2\mathcal{L}_{G}\big)\Big) \ .
\label{eq:GR_stress_energy_tensor}
\end{equation}
If for a given scalar field with Lagrangian $F(X,\phi)$, with $X\equiv g^{\mu\nu}\partial_\mu\phi\partial_\nu\phi$, minimally coupled to $g_{\mu\nu}$, it is possible to find a new scalar field Lagrangian $K(Z,\phi)$, with $Z\equiv h^{\mu\nu}\partial_\mu\phi\partial_\nu\phi$, minimally coupled to $h_{\mu\nu}$, such that Eq.~(\ref{eq:GR_stress_energy_tensor}) is satisfied, then our goal of mapping the dynamics of $f(R,Q)$ theories into GR coupled to a scalar field will be accomplished.

But before entering into technicalities, it is important to note that objects such as $\hat P$ and $\hat{\Omega}$, which by virtue of Eq.~(\ref{eq:square_completed}) are functions of the matter sources, can be expressed in two different ways, namely in terms of the original frame variables ${T^\mu}_\nu$, which are minimally coupled to $g_{\mu\nu}$; or in terms of the Einstein-frame variables  ${\tilde{T}^\mu}_{\ \ \nu}$, which are minimally coupled to $h_{\mu\nu}$. This basic property can be used to extract useful information about the structure of the theory and the relation between the $g_{\mu\nu}$ and the $h_{\mu\nu}$ (Einstein) frames, as we show next. \\

The stress-energy tensor ${T^{\mu}}_{\nu}$ that follows from the Lagrangian $F(X,\phi)$ (coupled to the $f(R,Q)$ theory) is obtained by varying the action
\begin{equation}
S_{F}= -\dfrac{1}{2}\int\text{d}^4x \sqrt{-g}F(X,\phi), 
\label{eq:matter_action_palatini}
\end{equation}
and takes the form 
\begin{equation}
{T^{\mu}}_{\nu}=\dfrac{-2g^{\mu\alpha}}{\sqrt{-g}}\dfrac{\delta S_F}{\delta g^{\alpha\nu}} = F_{X}{X^\mu}_\nu-\dfrac{F}{2}{\delta^\mu}_\nu ,
\label{eq:stress_energy_P}
\end{equation}
where $F_{X}\equiv \tfrac{\partial F}{\partial X}$ and ${X^\mu}_\nu\equiv g^{\mu\alpha}\partial_\alpha\phi\partial_\nu\phi$. 
It is important to notice that the matrix $\hat X$ with components ${X^\mu}_\nu$ satisfies the relation 
\begin{equation}
\label{eq:property_Xmn_gen}
\hat X^n = X^{n-1} \hat X \ ,
\end{equation}
where $X\equiv g^{\mu\nu}\partial_\mu\phi\partial_\nu\phi$ is the trace of $\hat X$. Because of this relation, any function of 
${T^{\mu}}_{\nu}$, such as $\hat \Omega$, can be formally written as a linear combination of the identity matrix and of $\hat X$: 
\begin{equation}\label{eq:OmfF}
{\Omega^\mu}_\nu=C \delta^\mu_\nu+D {X^\mu}_\nu \ ,
\end{equation}
where $C$ and $D$ are some functions of the scalars $X$ and $\phi$. 

 Analogously, we assume that there exists a scalar field Lagrangian $K(Z,\phi)$ (coupled to GR) such that 
\begin{equation}
{\tilde{T}^{\mu}}_{\ \ \nu}=\dfrac{-2h^{\mu\alpha}}{\sqrt{-h}}\dfrac{\delta S_K}{\delta h^{\alpha\nu}} =K_{Z}{Z^\mu}_\nu-\dfrac{K}{2}{\delta^\mu}_\nu \, ,
\label{eq:stress_energy_K}
\end{equation}
where $K_{Z}\equiv \tfrac{\partial K}{\partial Z}$, ${Z^\mu}_\nu\equiv h^{\mu\alpha}\partial_\alpha\phi\partial_\nu\phi$, and $Z={Z^\mu}_\mu$. Since ${Z^\mu}_\nu$ satisfies a relation identical to Eq.~(\ref{eq:property_Xmn_gen}), it follows that we can also write $\hat \Omega$ as 
\begin{equation}\label{eq:OmEF}
{\Omega^\mu}_\nu=\tilde{C} \delta^\mu_\nu+\tilde{D} {Z^\mu}_\nu \ ,
\end{equation}
where $\tilde{C}$ and $\tilde{D}$ are functions of $Z$ and $\phi$. \\

With the formal expressions (\ref{eq:OmfF}) and  (\ref{eq:OmEF}), we can now explore the relations between the scalars $R={P^\mu}_\mu=g^{\mu\nu}R_{\mu\nu}(h)$ and $\mathcal{R}\equiv h^{\mu\nu}R_{\mu\nu}(h)$. From Eq.~(\ref{eq:Einstein_frame_GR}), it is easy to see that $R_{\mu\nu}(h)=\kappa^2(K_Z Z_{\mu\nu}-h_{\mu\nu}(ZK_Z-K)/2)$. For simplicity, we will focus on the particular case of a free scalar field, namely, $K(Z,\phi)=Z$, for which $R_{\mu\nu}(h)=\kappa^2 Z_{\mu\nu}$ and $\mathcal{R}=\kappa^2 Z$ take their simplest forms. Now, given that $R=g^{\mu\nu}R_{\mu\nu}(h)$ and that $g^{\mu\nu}=\tilde{C}h^{\mu\nu}+\tilde{D}Z^{\mu\nu}$ (recall Eq.~\eqref{eq:metric_relation1}), we find that 
\begin{equation}\label{eq:R1}
R=\kappa^2 Z (\tilde{C}+Z \tilde{D}) \ . 
\end{equation}
Similarly, we can evaluate the relation between ${X^\mu}_\nu$ and ${Z^\mu}_\nu$, which takes the form ${X^\mu}_\nu=(\tilde{C}+Z \tilde{D}){Z^\mu}_\nu$ and leads to $X=Z (\tilde{C}+Z \tilde{D})$. By direct comparison of this result with Eq.~(\ref{eq:R1}), we see that, regardless of the gravity Lagrangian $f(R,Q)$, if $K(Z,\phi)=Z$, we have that $R=\kappa^2 X$. A similar analysis for the scalar $Q$ shows that in this case we also have $Q=(\kappa^2 X)^2=R^2$, for any $f(R,Q)$ theory. \\

\subsection{Matter Lagrangian $F(X,\phi)$ when $K(Z,\phi)=Z$} 
Substituting Eqs.~\eqref{eq:stress_energy_P} and~\eqref{eq:stress_energy_K} into Eq.~\eqref{eq:GR_stress_energy_tensor} and comparing the left- and right-hand sides of Eq.~\eqref{eq:GR_stress_energy_tensor}; taking also into account the scalar field equation in the two frames \cite{Afonso:2018hyj}, we obtain the relations 
\begin{align}
\label{eq:map_equations_1} K_{Z}{Z^\mu}_\nu&= \dfrac{F_{X}{X^\mu}_\nu}{{|\hat{\Omega}|}^{1/2}},\\
\label{eq:map_equations_2} K&= \dfrac{2\mathcal{L}_{G}+F_{X}X-F}{{|\hat{\Omega}|}^{1/2}}.
\end{align}
Using the trace of Eq.~\eqref{eq:map_equations_1} in Eq.~\eqref{eq:map_equations_2}, we can find an expression for the Lagrangian $F(X,\phi)$, as a function of $Z$ and $\phi$, given by 
\begin{equation}\label{eq:matter_lagrangian_P_map0}
F(X,\phi) = 2\mathcal{L}_{G} + \sqrt{|\hat{\Omega}|}\left[ZK_{Z}-K\right] \ .
\end{equation}
It is thus evident from this general expression that the task of finding the modified matter Lagrangian $F(X,\phi)$, when a certain $K(Z,\phi)$ has been specified, cannot be accomplished until a specific gravity Lagrangian $\mathcal{L}_{G}(R,Q)$ is provided, which also determines the specific form of $\hat\Omega$. In particular, we need the expressions for  $X=X(Z,\phi)$, $R=R(X,\phi)$, and $Q=Q(X,\phi)$; in order to express $F(X,\phi)$ explicitly (or parametrically) as a function of $X$ (or $Z$) and $\phi$. However, in the particular case of $K=Z$, an important simplification occurs, namely, 
\begin{equation}
F(X) = 2\mathcal{L}_{G} \, ,
\label{eq:matter_lagrangian_P_map}
\end{equation}
such that the form of the Lagrangian $F=F(X)$ is completely specified by the gravity Lagrangian $\mathcal{L}_{G} =f(R,Q)/2\kappa^2$. In addition, given that for $K=Z$ we always have $R=\kappa^2 X$ and $Q=(\kappa^2 X)^2$, it follows that the modified matter Lagrangian is given by 
\begin{equation}
F(X) = \frac{1}{\kappa^2}f(\kappa^2 X,(\kappa^2 X)^2) \ .
\label{eq:matter_lagrangian_P_map}
\end{equation}
This is a key result that will be used thoroughly during the rest of the paper. Note that Eq.~\eqref{eq:matter_lagrangian_P_map} implies that the nonlinearity of the gravitational sector is inherited by the matter sector via the mapping process. 

\subsection{Deformation matrix $\hat \Omega$ when $K(Z,\phi)=Z$}
In order to find the deformation matrix $\hat \Omega$, one must first find the explicit dependence of $\hat P$ on the matter sources. For this purpose, we rewrite the components of the left-hand side of Eq.~\eqref{eq:square_completed} as ${M^\mu}_{\alpha}{M^\alpha}_{\nu}$, where
\begin{equation}
{M^\mu}_\nu= \left({P^\mu}_\nu+\dfrac{f_{R}}{4f_{Q}}{\delta^\mu}_\nu\right),
\label{eq:M_tensor_definition}
\end{equation} 
and given that {(by the Cayley-Hamilton theorem)} any function of $\hat P$, like $\hat \Omega$ in Eq.~(\ref{eq:OmfF}), can be written as a linear combination of the identity and $\hat X$, we must have 
\begin{equation}
{M^\mu}_\nu\equiv  \beta{\delta^\mu}_\nu + \gamma {X^\mu}_\nu \ .
\label{eq:M_tensor_equation}
\end{equation}  
Inserting this expression in Eq.~\eqref{eq:square_completed} and identifying terms, we find that 
\begin{eqnarray}
\beta^2&=& \left(\frac{f_R}{4f_Q}\right)^2+\frac{f}{4f_Q}-\frac{\kappa^2 F(X,\phi)}{4f_Q} \ , \\
0&=& X \gamma^2+2\beta \gamma-\frac{\kappa^2 F_X}{2f_Q} \ .
\label{gamma}
\end{eqnarray}
Equation~\eqref{gamma} defines the function $\gamma$ in terms of a second-order algebraic equation. But before solving it, it is convenient to focus in our case of interest $K(Z)=Z$, which leads to Eq.~\eqref{eq:matter_lagrangian_P_map}. Given that $\mathcal{L}_{G}=f(R,Q)/2\kappa^2$, Eq.~\eqref{eq:matter_lagrangian_P_map} implies that $F(X)=f(R,Q)/\kappa^2$ and leads to $\beta^2=\left(\frac{f_R}{4f_Q}\right)^2$. One can check that in order to recover the correct $f(R)$ limit when $f_Q\to 0$, only the positive sign is allowed when solving for $\beta$, thus giving $\beta=f_R/4f_Q$. As a result, and demanding again the correct limit when $f_Q\to 0$, one finds that 
\begin{equation}\label{eq:gamma}
\gamma= \frac{-f_R+\sqrt{f_R^2+8f_Q \kappa^2 X F_X}}{4f_Q X} \ .
\end{equation}
With the results for $\beta$ and $\gamma$ just obtained, we can use Eq.~\eqref{eq:M_tensor_definition} to finally write $\hat P$, when $K(Z)=Z$, as
\begin{equation}\label{eq:hatP}
\hat P= \gamma \hat X \ .
\end{equation}
A consequence of this relation is that $R={P^\mu}_\mu=\gamma X$ should be consistent with our previous result $R=\kappa^2X$, which is not obvious, given the form of Eq.~\eqref{eq:gamma}. However, using Eq.~(\ref{eq:matter_lagrangian_P_map}) one can see that 
\begin{equation}\label{eq:FX}
F_X=f_R+2\kappa^2 X f_Q \ ,
\end{equation}
which, when inserted in Eq.~(\ref{eq:gamma}), allows us to obtain 
\begin{equation}\label{eq:gamma2}
\gamma= \frac{-f_R+\sqrt{(f_R+4f_Q \kappa^2 X )^2}}{4f_Q X}=\kappa^2 \ ,
\end{equation}
and thus confirming the consistency of this with our previous results. Equation~\eqref{eq:hatP} can be written simply as 
\begin{equation}\label{eq:hatP2}
\hat P= \kappa^2 \hat X \ .
\end{equation}

Continuing with our search for $\hat \Omega$, let us write explicitly the form of $\hat \Omega^{-1}$, using Eqs.~(\ref{eq:Omega_sigma}) and (\ref{eq:sigma_tensor}), as 
\begin{equation}
{\left[\hat{\Omega}^{-1}\right]}^\mu_{\ \ \nu}= \frac{1}{|\hat{\Omega}|^{1/2}}\left(f_R\delta^\mu_\nu+2f_Q \kappa^2 {X^\mu}_\nu\right) \ ,
\end{equation}
from which it follows that $|\hat{\Omega}|^{1/2}=|f_R\delta^\mu_\nu+2f_Q \kappa^2 {X^\mu}_\nu|$. Evaluating this quantity and manipulating the result, one obtains 
\begin{equation}\label{eq:detOm}
|\hat{\Omega}|={f_R^3}\left(f_R+2f_Q \kappa^2 X\right) \, .
\end{equation}
Writing $\hat \Omega^{-1}$ as $\hat \Omega^{-1}=A \hat I+B\hat X$, it follows that $\hat \Omega=C \hat I+D\hat X$, with 
\begin{equation}
C=\frac{1}{A} \ , \ D=-\frac{B}{A}\frac{1}{(A+BX)} \ ,
\end{equation}
which explicitly reads
\begin{equation}\label{eq:OmGeneral}
{\Omega}^\mu_{\ \ \nu}= \sqrt{f_R(f_R+2\kappa^2X f_Q)}\left(\delta^\mu_\nu-\frac{2\kappa^2f_Q }{f_R+2\kappa^2X f_Q} {X^\mu}_\nu\right) \ .
\end{equation}

Now one can use the above expression for $\hat \Omega$ and its inverse to find the relations between $g_{\mu\nu}$ and $h_{\mu\nu}$ and their inverses. In particular, from Eq.~(\ref{eq:metric_relation1}) it is easy to see that $g_{\mu\nu}=h_{\mu\alpha}{[\Omega^{-1}]}^\alpha_{\ \ \nu}$ can be written as
\begin{equation}\label{eq:gh}
g_{\mu\nu}=\frac{f_R}{|\hat{\Omega}|^{1/2}}h_{\mu\nu}+\frac{2\kappa^2f_Q}{F_X }\partial_\mu \phi \partial_\nu \phi  \ .
\end{equation} 
To obtain this result, we used that $h_{\mu\alpha}{Z^\alpha}_\nu=\partial_\mu \phi \partial_\nu \phi$ and the relation ${Z^\alpha}_\nu=F_X {X^\alpha}_\nu/|\hat{\Omega}|^{1/2} $, that follows from Eq.~(\ref{eq:map_equations_1}), when $K=Z$. Tracing this quantity one also finds a relation between the scalars $Z$ and $X$, of the form
\begin{equation}\label{eq:ZofX}
Z=\frac{X F_X}{|\hat{\Omega}|^{1/2}} \ .
\end{equation}
If this expression can be inverted to obtain $X=X(Z)$, then all the scalar quantities appearing in Eq.~(\ref{eq:gh}) can be written in terms of solutions obtained in the Einstein frame theory. In the following we will use that property to build new solutions in $f(R,Q)$ theories generated by a free massless scalar in GR. 

\section{Quadratic $f(R,Q)$ model}\label{sec:fRQQuad}
Now that we have constructed the map between GR and $f(R,Q)$ gravity, we can apply this algorithm to find (non-trivial) scalar configurations for a given $f(R,Q)$ gravity model. 
Let us concentrate our attention in mapping scalar solutions of GR into solutions of the quadratic gravity model
\begin{equation}
f(R,Q) = R + a\,R^2+b\,Q,
\label{eq:f_model}
\end{equation}
that reduces to GR if the free parameters, $a$ and $b$, vanish. The derivatives of $f(R,Q)$ are $f_{R}=1+2a\,R$ and $f_{Q}=b$.

\begin{figure*}[!h]
\centering
\includegraphics[width=\columnwidth]{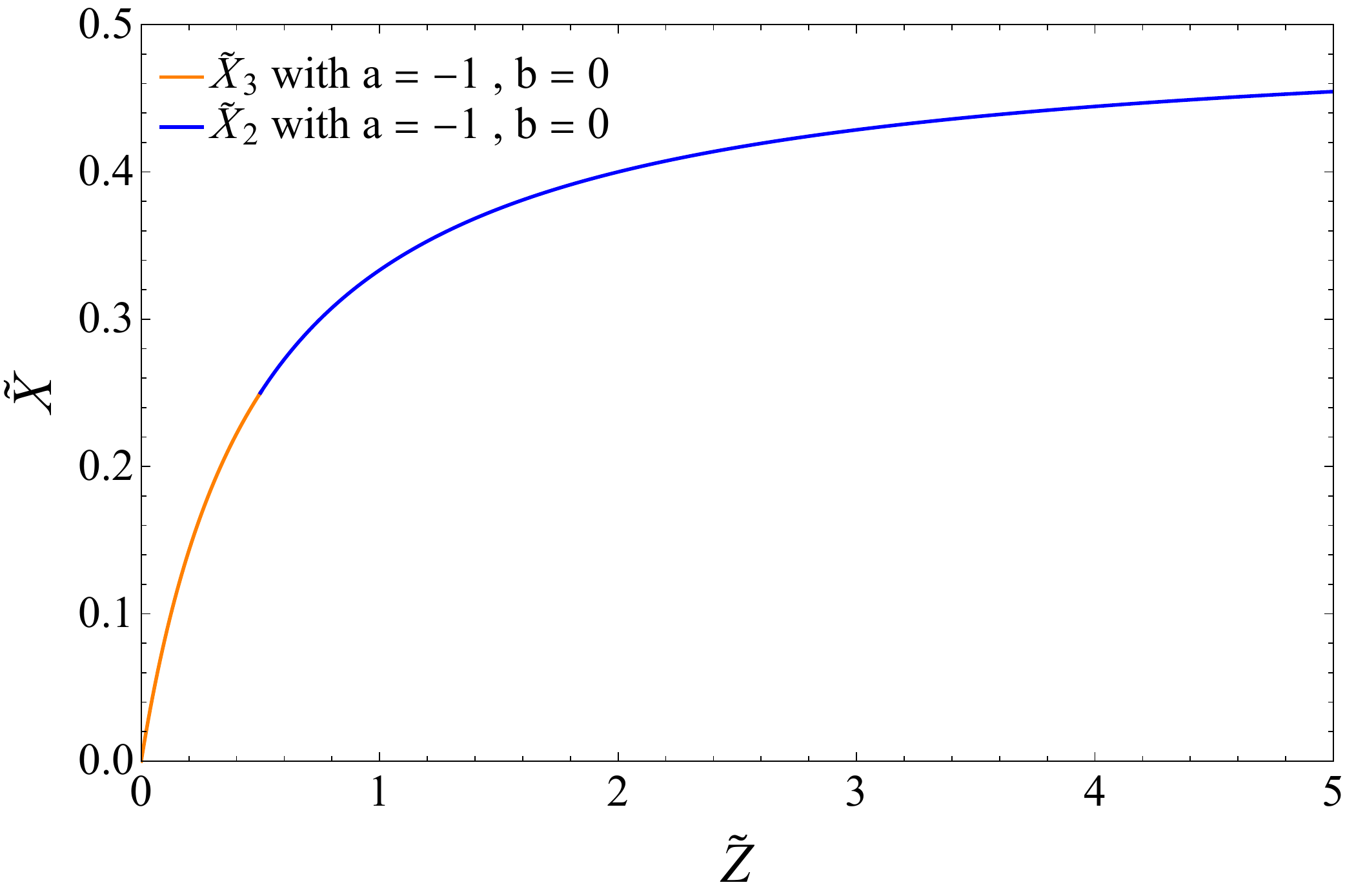} \includegraphics[width=\columnwidth]{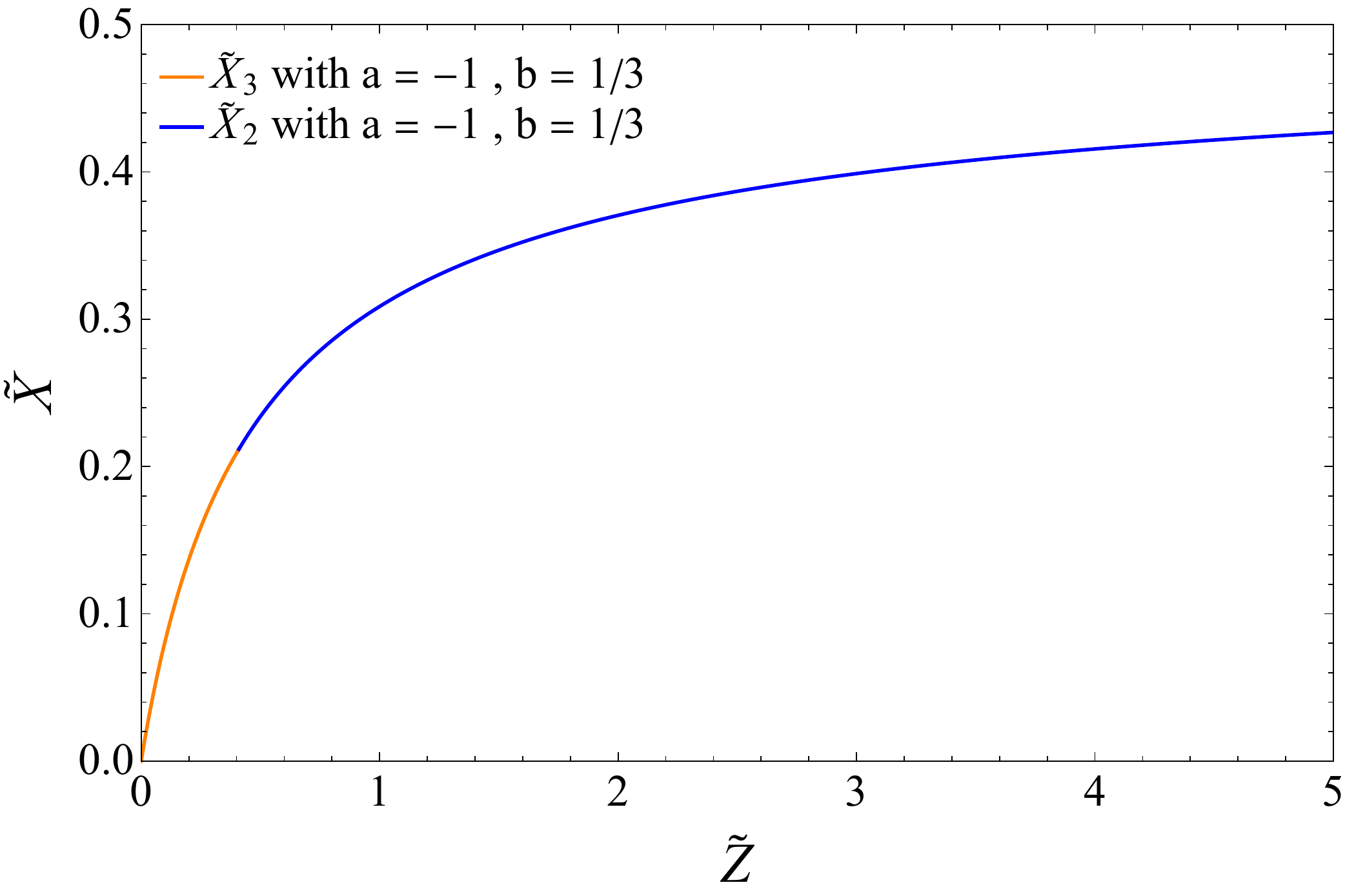}
\caption{When $a+b<0$, with $b\geq 0$, the function $\tilde{X}(\tilde{Z})$ tends smoothly to a maximum
asymptotic value given by $\tilde{X}_{\infty}=-1/(2a)$, as $\tilde{Z}\rightarrow\infty$. In the left panel we show this behavior for $a=-1$ and $b=0$; while in the right panel we plot $\tilde{X}$ for $a=-1$ and $b=1/3$.}
\label{fig:xz1}
\end{figure*}

\begin{figure*}[!h]
\centering
\includegraphics[width=\columnwidth]{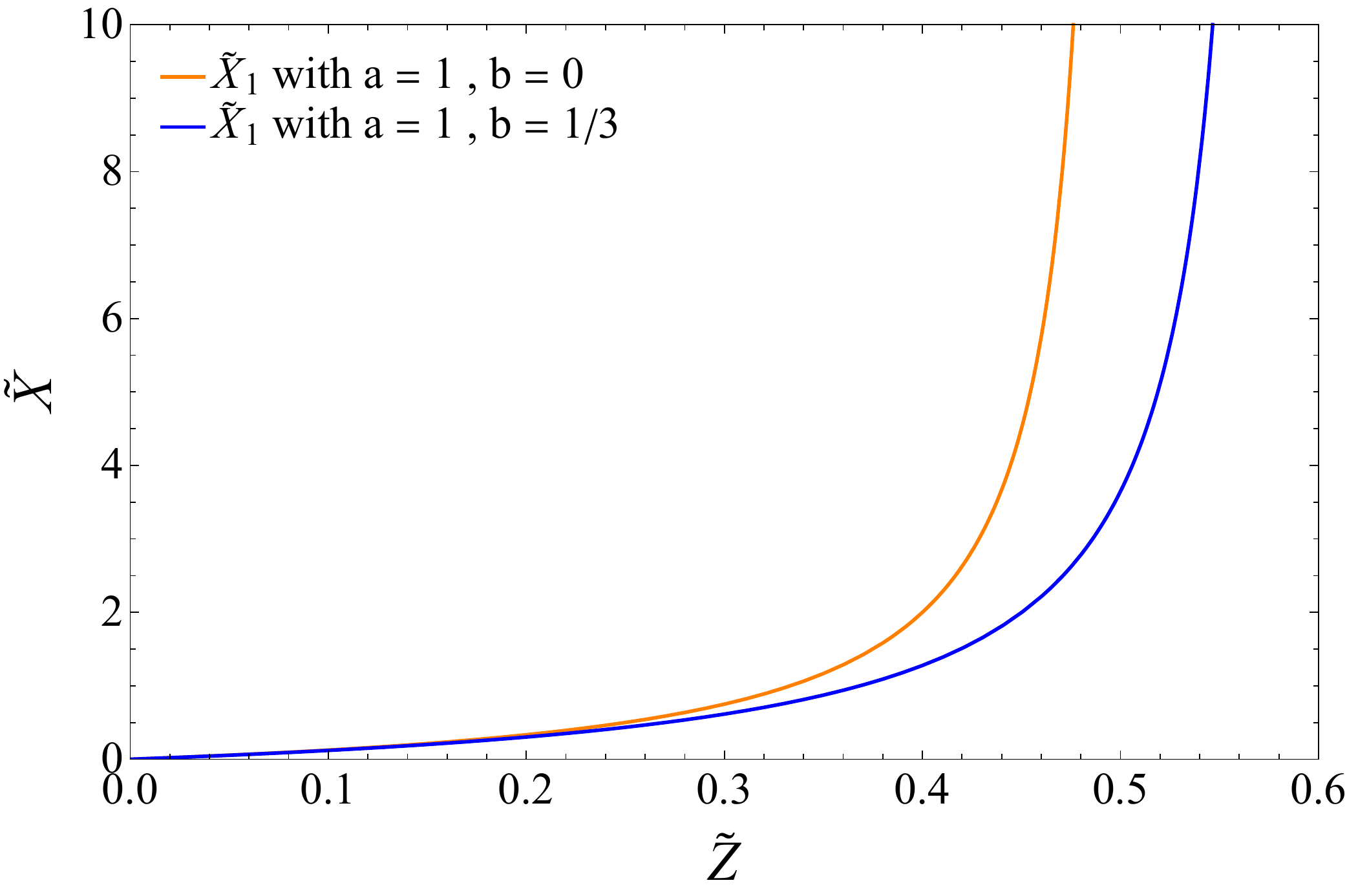} \includegraphics[width=\columnwidth]{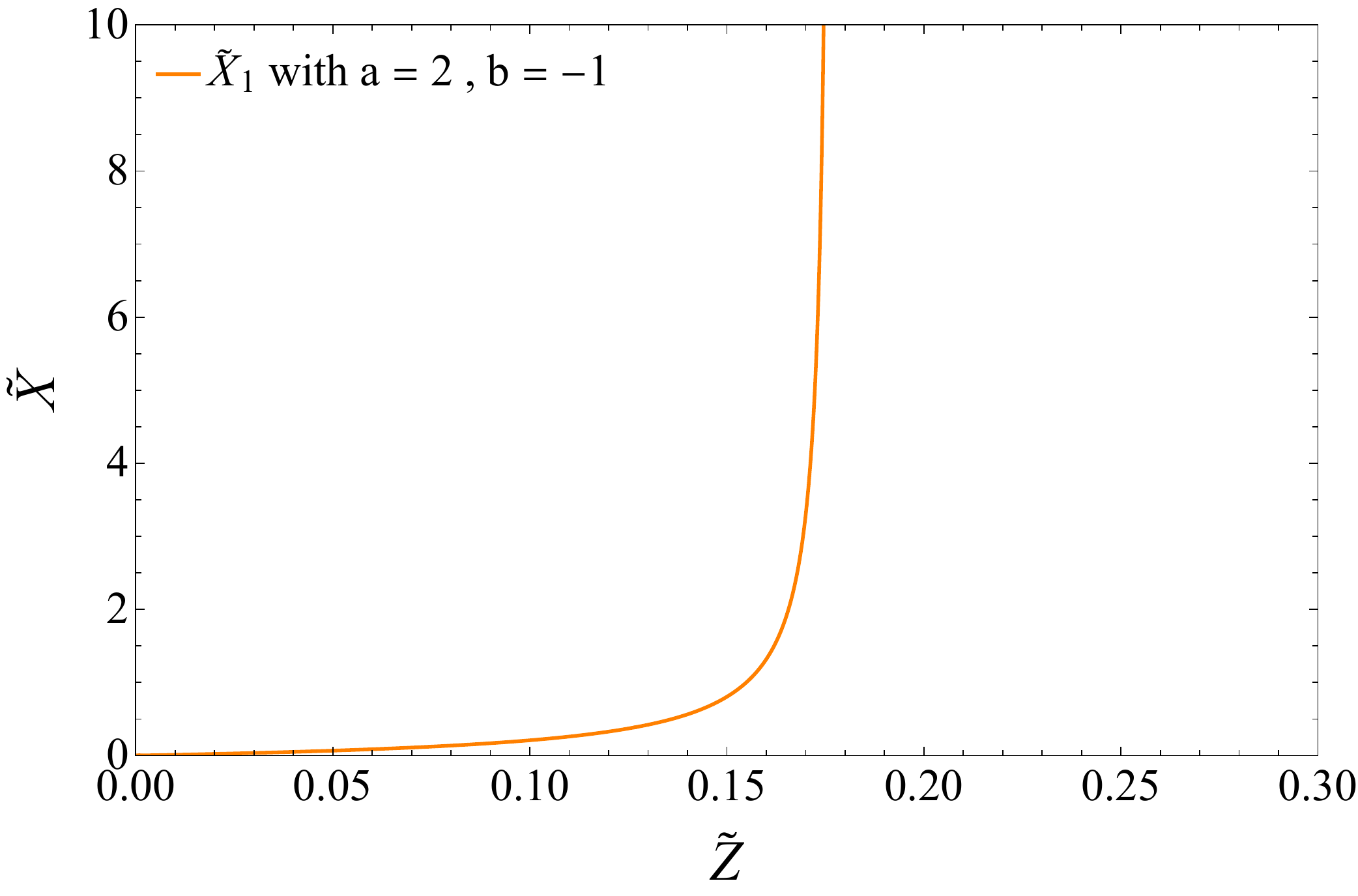}
\caption{When $a+b>0$, with $a>-b$, the function $\tilde{X}(\tilde{Z})$ blows up at a critical value of $\tilde{Z}$, given by $\tilde{Z}_c = \sqrt{(a+b)/4a^3}$. In the left panel we show this behavior for $a=1$ and non-negative values of $b$, namely $b=0$ and $b=1/3$; while in the right panel we plot $\tilde{X}$ for $a=2$ and $b=-1$.}
\label{fig:xz2}
\end{figure*}

\begin{figure*}[!h]
\centering
\includegraphics[width=\columnwidth]{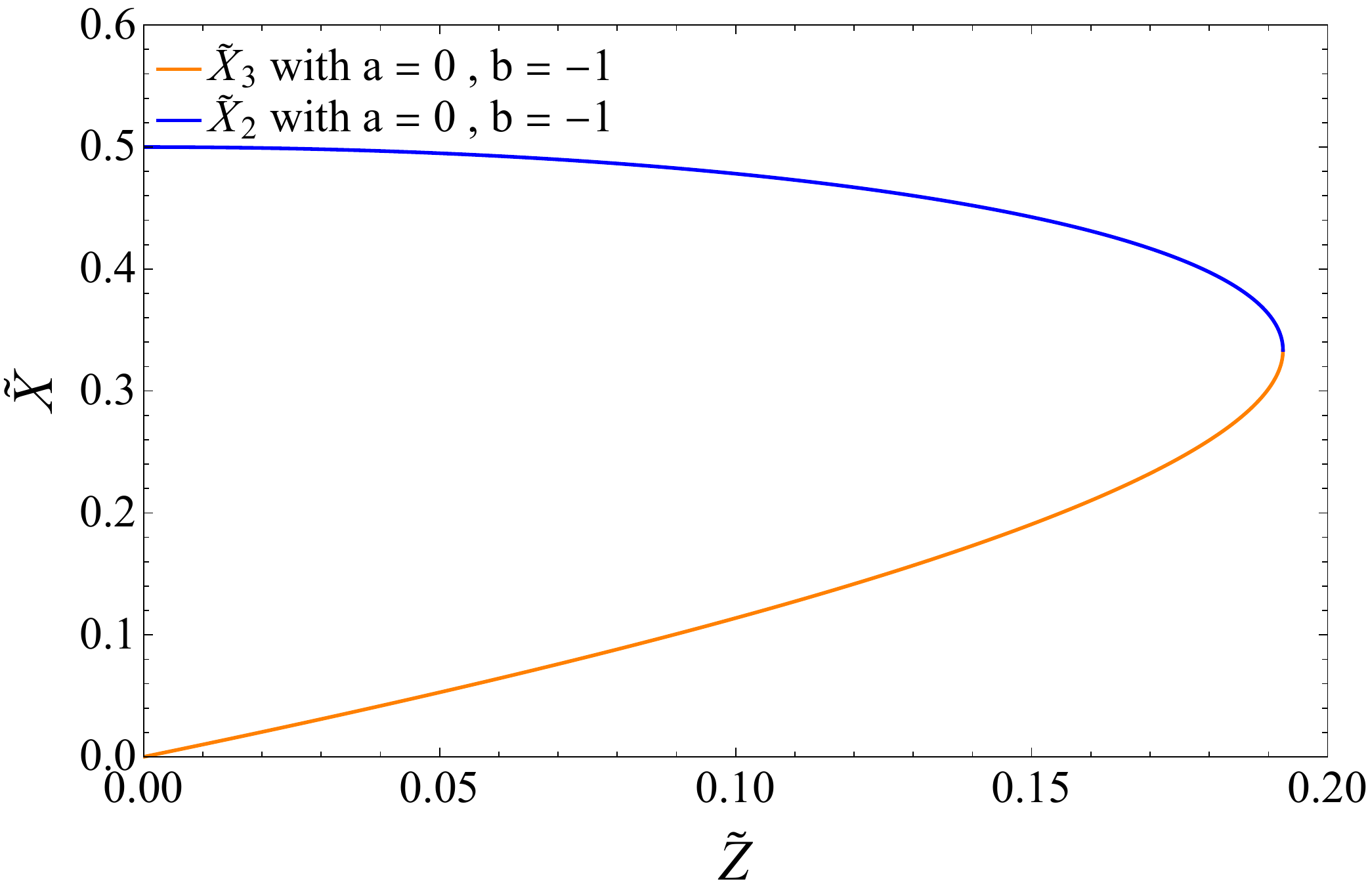} \includegraphics[width=\columnwidth]{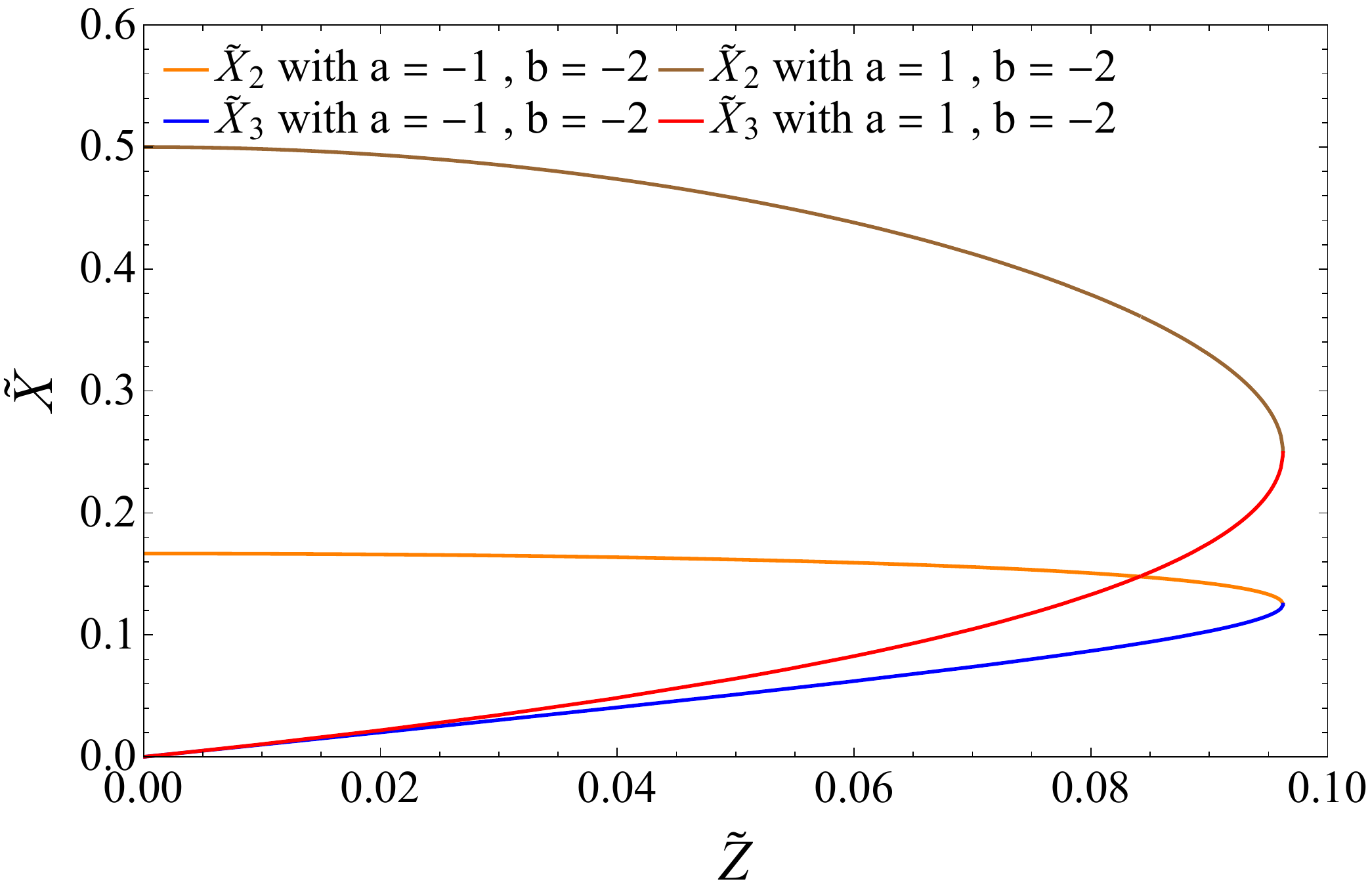}
\caption{When $a+b<0$, with $b<0$, the function $\tilde{X}(\tilde{Z})$ is not single
valued in its interval of definition. In the left panel we show this behavior for $b=-1$ and $a=0$; and in the right panel we set $b=-2$ and choose two values of $a$, namely $a=1$ and $a=-1$.}
\label{fig:xz3}
\end{figure*}

\begin{figure}[!h]
\centering
\includegraphics[width=\columnwidth]{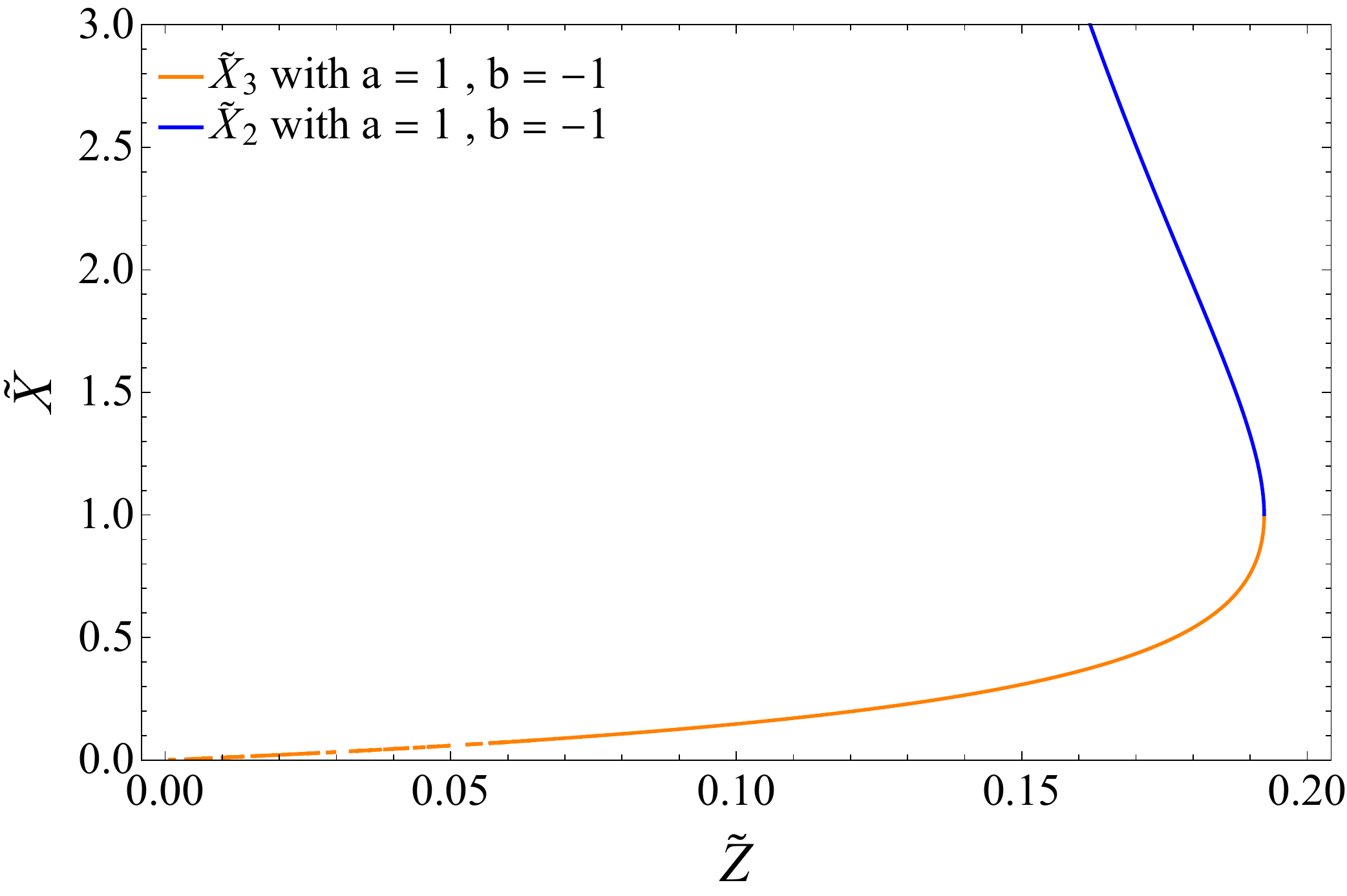} \includegraphics[width=\columnwidth]{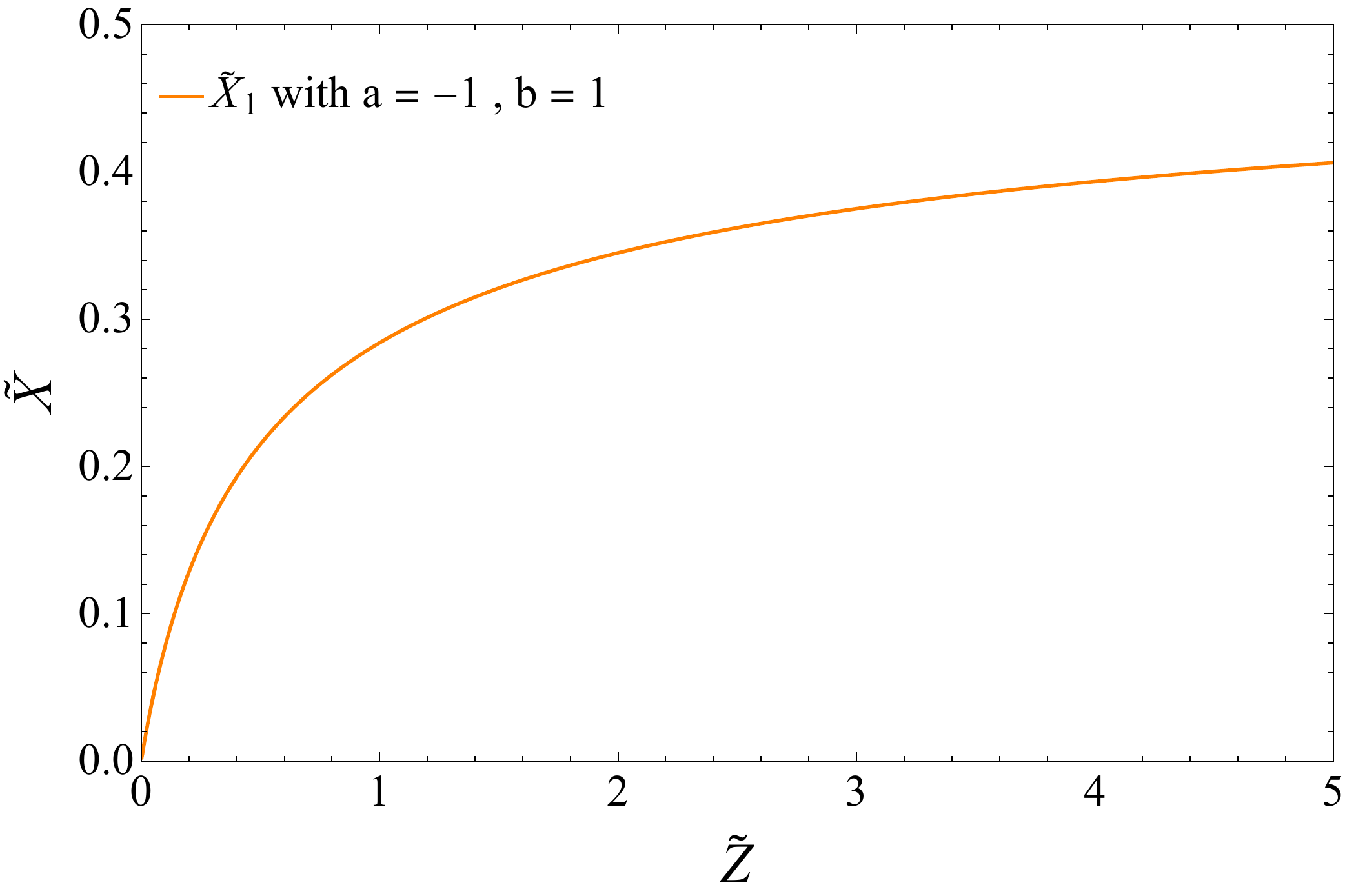}
\caption{When $a+b=0$ the function $\tilde{X}(\tilde{Z})$ has two different behaviors, depending on the sign choice of the parameters $a$ and $b$. If $a>0$ and $b<0$, $\tilde{X}(\tilde{Z})$ is not a single valued function along its interval of definition, so one must assume that only the lower branch is physically interesting (we show this behavior in the top panel for $a=1$ and $b=-1$); whereas if $a<0$ and $b>0$, the function $\tilde{X}(\tilde{Z})$ is limited, and tends smoothly to a maximum asymptotic value $\tilde{X}_{\infty}=-1/(2a)$ (we show this behavior in the bottom panel for $a=-1$ and $b=1$).}
\label{fig:xz4}
\end{figure}

\begin{figure}[ht]
\centering
\includegraphics[width=\columnwidth]{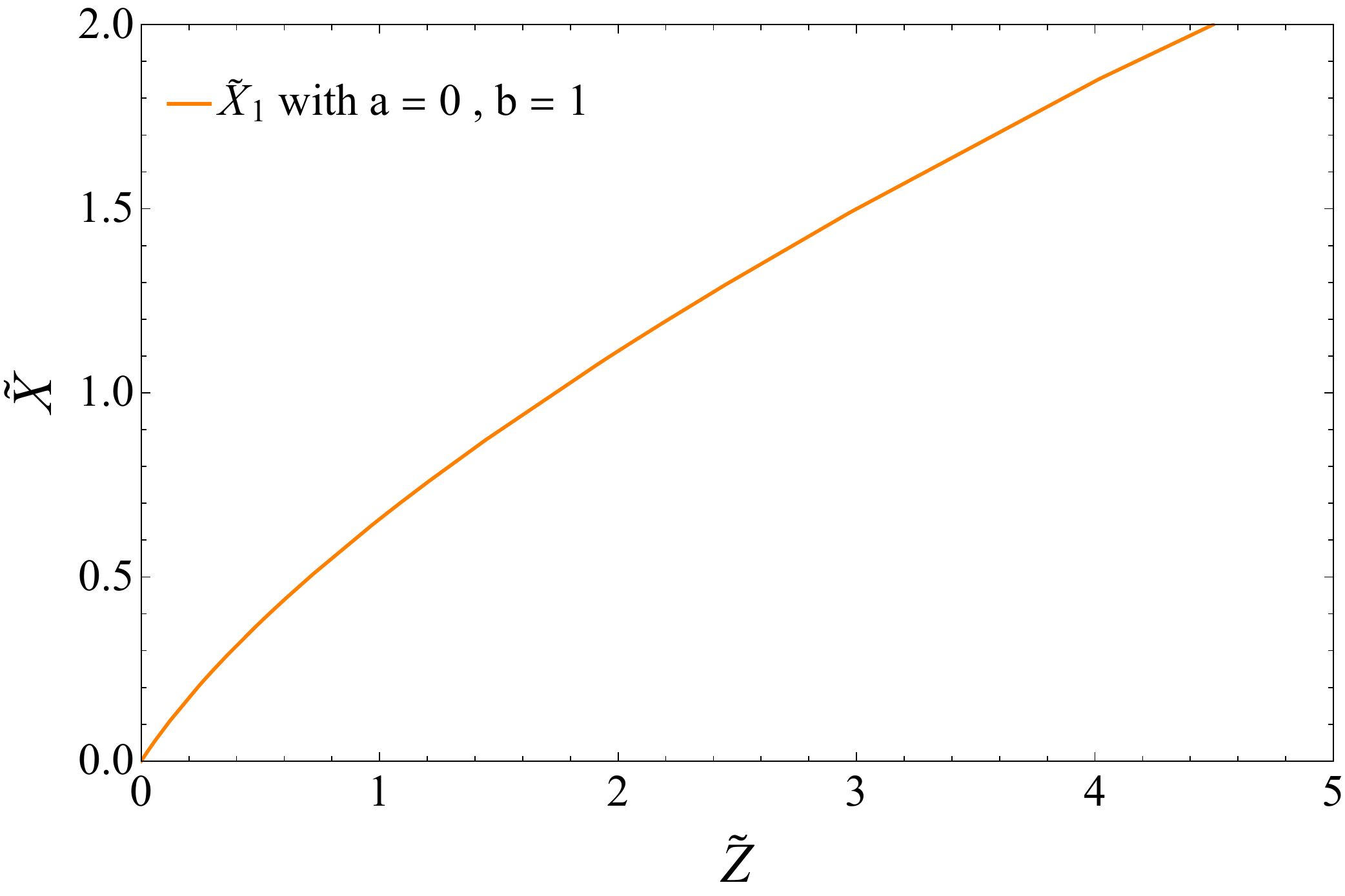}
\caption{When $a+b>0$ and $a=0$, the function $\tilde{X}(\tilde{Z})$ grows indefinitely as $\tilde{Z}$ increases. For very large values of $\tilde{Z}$ the function $\tilde{X}(\tilde{Z})$ has the asymptotic behaviour $\tilde{X}(\tilde{Z})\approx (\tilde{Z}^2/(2b))^{1/3}$. For this plot we have chosen $a=0$ and $b=1$.}
\label{fig:xz5}
\end{figure}

In order to generate new scalar configurations using the map 
described in Sec.~\ref{c_map}, let us consider GR minimally coupled to a scalar field, with matter action
\begin{equation}
\label{eq:matter_action_Z_GR}
\tilde{S}_{\phi}(Z)=-\dfrac{1}{2}\int \text{d}^4x\sqrt{-h}Z=-\dfrac{1}{2}\int \text{d}^4x\sqrt{-h}h^{\mu\nu}\partial_{\mu}\,\phi\partial_{\nu}\,\phi,
\end{equation}
so that we have $K(Z,\phi)=Z$. In this scenario, the matter Lagrangian coupled to the quadratic model~\eqref{eq:f_model}, given by Eq.~\eqref{eq:matter_lagrangian_P_map}, reduces to
\begin{equation}
F(X) = \dfrac{f(R,Q)}{\kappa^2}= X+(a+b)\,\kappa^2X^2 \ ,
\label{eq:matter_lagramgian_F_map_model}
\end{equation}
and the relation between metrics is given by 
\begin{equation}\label{eq:ghQuad}
g_{\mu\nu}=\frac{1}{(f_RF_X)^{1/2}}h_{\mu\nu}+\frac{2\kappa^2 b}{F_X}\partial_\mu \phi \partial_\nu \phi  \ ,
\end{equation} 
with
\begin{eqnarray}
f_R&=&1+2a \kappa^2 X \nonumber \, ,  \\
F_X&=&1+2(a+b) \kappa^2 X \ . \label{eq:fRfX}
\end{eqnarray}
Since we are interested in using the GR solutions as seed for new solutions in the $f(R,Q)$ theory, we need to find the explicit relation between $X$ and $Z$, which is given by 
\begin{equation}
Z=\frac{X F_X}{|\hat{\Omega}|^{1/2}}=X\sqrt{\frac{1+2(a+b) \kappa^2 X}{(1+2a \kappa^2 X)^3} }\ .
\end{equation}
Multiplying this equation by $\kappa^2$, squaring it, and defining $\tilde{Z}\equiv \kappa^2 Z$ and $\tilde{X}\equiv\kappa^2 X$, we find a cubic polynomial of the form
\begin{equation}
0= \alpha_3 \tilde{X}^3+\alpha_2 \tilde{X}^2+\alpha_1 \tilde{X}+\alpha_0  \ ,
\label{3_deg_poly}
\end{equation}
where
\begin{eqnarray}
\alpha_3&=&  \left(8 a^3 \tilde{Z}^2-2 (a+b)\right) \, , \\
\alpha_2&=& \left(12 a^2 \tilde{Z}^2-1\right) \, , \\
\alpha_1&=&6 a \tilde{Z}^2 \, , \\
\alpha_0&=&\tilde{Z}^2 \ .
\end{eqnarray}
This type of equations have three different solutions which are tabulated and well known, but have an awkward form that we will not write explicitly. For simplicity we will refer to them as $\{\tilde{X}_1,\tilde{X}_2,\tilde{X}_3\}$. The criterion to identify the physical solutions is based on the small $Z$ limit, which should lead to $\tilde{X}\approx \tilde{Z}$,  and on the continuity and differentiability of the resulting curve at higher values of $\tilde{Z}$. This allows us to split the solutions in two big groups, depending on whether the combination $a+b=0$ or $a+b\neq 0$. In the latter case, it is also important to distinguish between $a+b>0$ and $a+b<0$. Finally, when $a+b<0$, we can have $a\cdot b>0$ (both negative) or $a\cdot b<0$ (only one negative). The case $a\cdot b=0$ must also be considered separately. The different behaviors of the resulting curves are shown in Figs.~\ref{fig:xz1},~\ref{fig:xz2},~\ref{fig:xz3},~\ref{fig:xz4} and~\ref{fig:xz5}. \\

From these plots we identify four types of behaviours in the function $\tilde{X}(\tilde{Z})$. The first one appears in Fig.~\ref{fig:xz1} and in the bottom panel of Fig.~\ref{fig:xz4}, and corresponds to a curve which asymptotically tends to a maximum value $\tilde{X}_{max}=-1/(2a)$. The second case is illustrated in  Fig.~\ref{fig:xz2}, representing an $\tilde{X}(\tilde{Z})$ with a divergence at $\tilde{Z}_c=\sqrt{(a+b)/4a^3}$. Then we have the case shown in Fig.~\ref{fig:xz3} and in the top panel of Fig.~\ref{fig:xz4}, in which $\tilde{X}(\tilde{Z})$ ceases to be single valued. Finally, we have an unbounded and monotonically growing $\tilde{X}(\tilde{Z})$, which is represented by just one of the three solutions of the third-degree polynomial \eqref{3_deg_poly} (see Fig.~\ref{fig:xz5}).

\section{Scalar configurations}\label{sec:sca_conf}
In this section we will explore the impact of the four types of solutions for $\tilde{X}(\tilde{Z})$ that we found previously, in spherically symmetric, static objects, engendered by a scalar field theory coupled to the quadratic $f(R,Q)$ theory (\ref{eq:f_model}). For this purpose we will use as seed the solution for this type of object that one finds in GR coupled to a scalar field characterized  by $K(Z,\phi)=Z$. With that solution at hand, we can use Eq.~\eqref{eq:ghQuad} to study the properties of the resulting geometries in the $f(R,Q)$ theory.  \\

\subsection{Static and spherically symmetric solution in GR}
\begin{figure}
\centering
\includegraphics[width=\columnwidth]{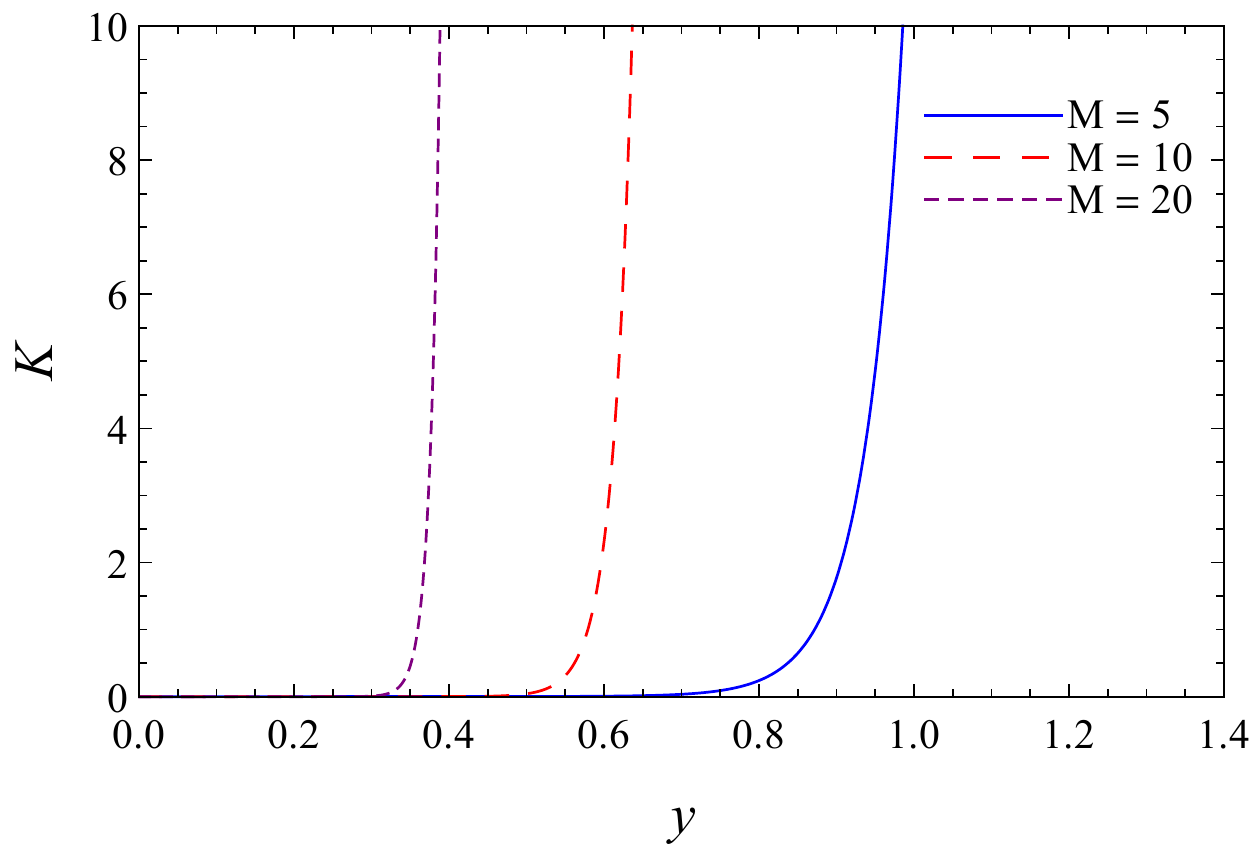}
\caption{The Kretschmann scalar of the JNW spacetime for different values of the ADM mass. In this plot we fixed $\mu^2=1.$}
\label{fig:KS}
\end{figure}

The spherically symmetric and static solution in GR coupled to the scalar field action \eqref{eq:matter_action_Z_GR} is well-known in the literature, usually called Janis-Newman-Winicour (JNW) spacetime~\cite{JNW:1968}. Under the Bergman and Leipnik assumptions~\cite{bergman}, Wyman showed that the integration of the field equations of GR coupled to Eq.~\eqref{eq:matter_action_Z_GR} is possible \cite{wyman}, which allowed him to find the JNW solution in a specific coordinate system
\begin{equation}
\label{eq:wyman}
ds^2_{GR} = -e^{\nu}\,dt^2 + \dfrac{e^{\nu}}{W^4}\,dy^2 + \dfrac{1}{W^2}\,d\sigma^2,
\end{equation}
where $d\sigma^2 = d\theta^2 + \sin^2\theta d\varphi^2$ is the line element of a unit sphere, and the functions $\nu$ and $W$ have only radial dependence (functions of the coordinate $y$). The line element~\eqref{eq:wyman} represents an asymptotically flat spacetime generated by a time-independent spherically symmetric scalar field distribution that satisfies the Klein-Gordon equation 
\begin{equation}
\label{eq:scalar_field_equation}
\Box \phi=0.
\end{equation}
In this coordinate system, the scalar field obeys the simple equation $\phi_{,yy}=0$~\cite{agnese}, and its solution is simply given by
\begin{equation}
\label{eq:field_distribution_wyman}
\phi(y)=\zeta\,y,
\end{equation}
where $\zeta$ is a constant that can be interpreted as the scalar charge. Requiring the line element~\eqref{eq:wyman} to be asymptotically flat, the $\nu$ and $W$ functions take the form
\begin{eqnarray}
\nu &=& -2M\,y, \nonumber \\
W&=& \dfrac{e^{-M\,y}\sinh(\eta\,y)}{\eta}, \label{eq:nuW}
\end{eqnarray}
where $M$ is the asymptotic ADM mass  
of the solution and $\eta \equiv \sqrt{M^2 + \mu^2/2}$, with $\mu^2\equiv \kappa^2\zeta^2$. In the $(t,y,\theta,\varphi)$ chart, the asymptotically flat region is reached as $y\to 0$, where the scalar field vanishes. This follows from the fact that the area of the spherical sector, $A=4\pi r^2$, is related to the $W$ function according to $r^2=1/W^2$. In the $y\to \infty $ limit it is easy to see that $W\approx e^{(\eta-M)y}/2\eta$, such that $r^2\approx (2\eta)^2e^{-2(\eta-M)y}$, which goes to zero because $\eta-M>0$. Instead, when $y\to 0$, we have $W\approx y$, leading to\footnote{Note that  $W$ and $y$ have dimensions of an inverse length, while $M,\eta$, and $\mu$  have dimensions of length.} 
 $r^2\approx 1/y^2$, which grows unboundedly. The region of highest energy density, therefore, corresponds to the limit $(y\to \infty)$, where the scalar field $\phi$ is singular. We then say that the spacetime has a physical singularity (see Fig.~\ref{fig:KS}). 
It is important to note that the line element~\eqref{eq:wyman} represents a horizonless object, because $g_{tt}$ never vanishes, and can thus be regarded as a naked singularity.

\subsection{Scalar configurations in $f(R,Q)$}
We already found in Eq.~\eqref{eq:ghQuad} the form of the metric generated by scalar configurations which in GR corresponds to a free massless field. We can now use the JNW solution specified by Eqs.~\eqref{eq:wyman},~\eqref{eq:field_distribution_wyman}, and~\eqref{eq:nuW}, together with the relation $X(Z)$ discussed in Sec.~\ref{sec:fRQQuad}, to explicitly evaluate the metric components of $g_{\mu\nu}$ in spherically symmetric and static scenarios. {This is what we need to explore the physics behind these solutions.} To proceed, we first write the line element associated to $g_{\mu\nu}$, which takes the form
\begin{equation}
ds^2=\frac{1}{(f_R F_X)^{1/2}}ds^2_{GR}+\frac{2\mu^2 b}{F_X}dy^2 \ .
\label{lel}
\end{equation}  
This line element has a piece that is conformal with the JNW geometry, but also has an additional (disformal) term coming from the $\partial_\mu\phi\partial_\nu\phi$ contribution. Since $\phi=\zeta y$, only the derivative in the $y$ direction provides a nonzero contribution, yielding the second term above (we recall that $\mu^2\equiv \kappa^2\zeta^2$). The line element ~\eqref{lel} can thus be rewritten as 
\begin{eqnarray}
ds^2&=&-\frac{e^\nu}{(f_R F_X)^{\frac{1}{2}}} dt^2+\left(\frac{e^\nu}{(f_R F_X)^{\frac{1}{2}}W^4} +\frac{2\mu^2 b}{F_X}\right)dy^2\nonumber \\ &+&\frac{1}{(f_R F_X)^{\frac{1}{2}}W^2}d\sigma^2 \ .
\end{eqnarray}  
Of particular interest is the radial function 
\begin{equation}\label{eq:r2}
r^2=\frac{1}{(f_R F_X)^{\frac{1}{2}}W^2} \ ,
\end{equation}
whose denominator now contains the functions $f_R$ and $F_X$ specified in Eq.~(\ref{eq:fRfX}). We will start our analysis by considering this quantity. 

\begin{figure}
\centering
\includegraphics[width=0.45\textwidth]{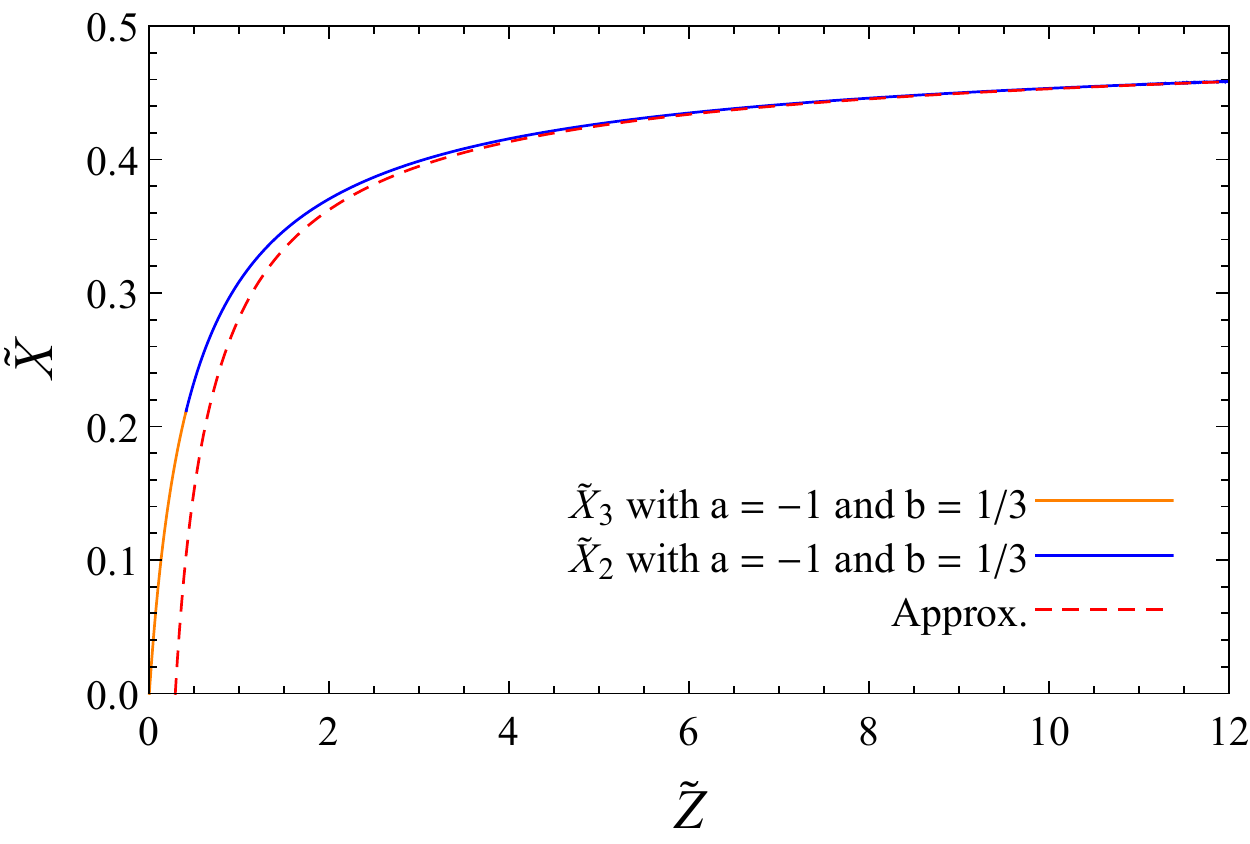}
\caption{Comparison of the solution for $\tilde{X}(\tilde{Z})$ shown in the right panel of Fig.~\ref{fig:xz1}, with the analytical approximation (\ref{eq:XZFit1}).}
\label{fig:GenAm1B1b3Fit}
\end{figure}

\begin{figure}
\centering
\includegraphics[width=0.45\textwidth]{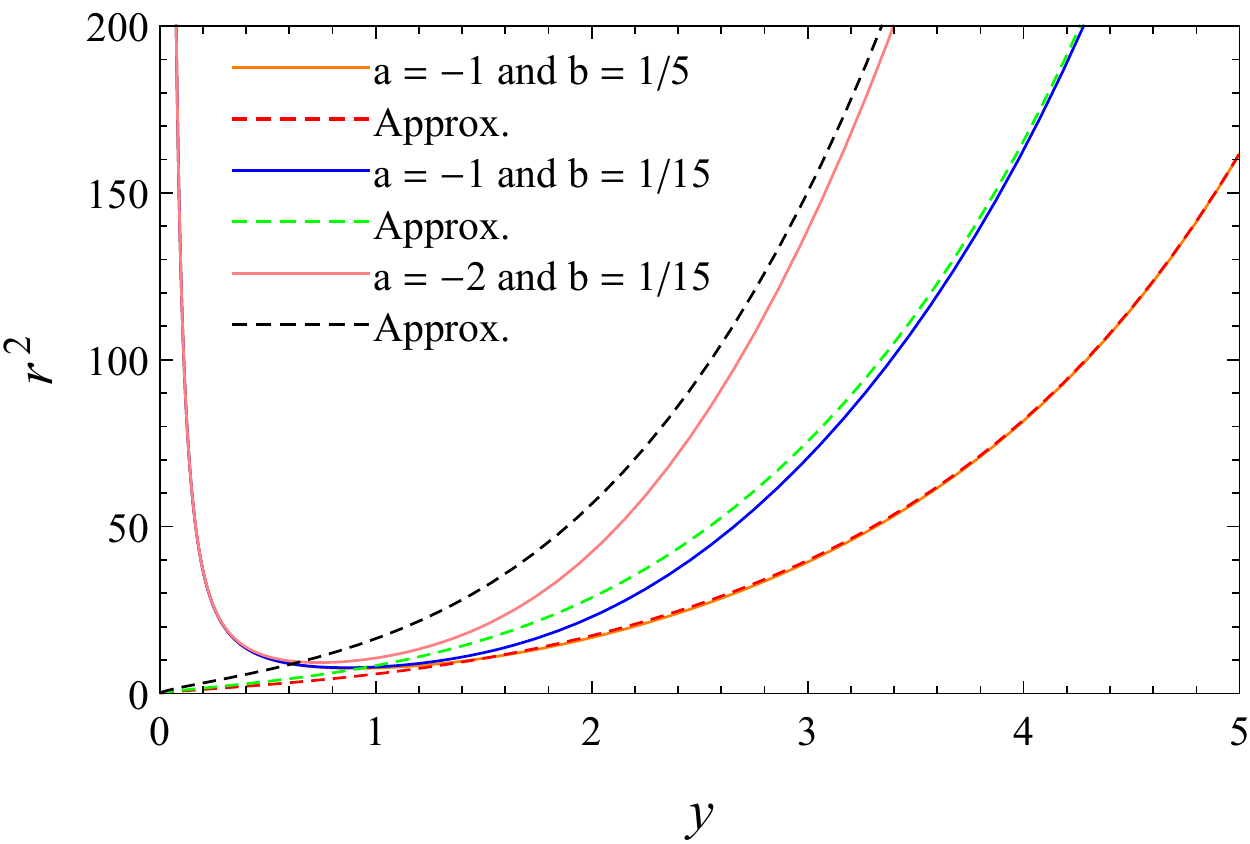}
\caption{Bounce in the area function $r^2(y)$, signaling the presence of a wormhole in this family of solutions. The analytical approximation for $y\to \infty$ is also shown (dashed curves). As $y\to 0$, the (solid) curves rapidly converge to the GR solution.}
\label{fig:R2WH}
\end{figure}

Let us begin by considering the behaviour of $r^2$ in the asymptotically flat region $y\to 0$. There we have $W\approx y$ and $\kappa^2 Z=\kappa^2 h^{\mu\nu}\partial_\mu \phi\partial_\nu\phi=\mu^2 W^4 e^{-\nu}\approx \mu^2 y^4\to 0$. Since in the limit $Z\to 0$ we must recover $X\approx Z$, one readily sees that $r^2$ and the whole geometry smoothly {recovers} the GR solution. On the other hand, given that in the central region, $y\to \infty$, we can approximate  $W\approx e^{(\eta-M)y}/2\eta$, it follows that $\kappa^2 Z=\mu^2 W^4 e^{-\nu}\approx \mu^2 e^{(4\eta-2M)y}/(2\eta)^4$ grows without bound in that region. Among the various behaviours that we observed in Sec.~\ref{sec:fRQQuad}, the case in which $X(Z)$ asymptotes to $\kappa^2 X_{max}=-1/(2a)$ as $Z\to \infty$ is particularly interesting.  A very good approximation for this scenario is given by 
\begin{equation}\label{eq:XZFit1}
\kappa^2X(Z)\approx -\frac{1}{2a }-\frac{\lambda}{(\kappa^2Z)^{2/3}} \ ,
\end{equation}
which in some cases requires considering the series expansion of $X_2$ and in others the expansion of $X_1$, depending on the specific parameters chosen. The specific form of the coefficient $\lambda$ is given by
\begin{equation}
\lambda= \frac{b^{1/3}}{a^2 2^{5/3}} \ .
\end{equation}
 A glance at Fig.~\ref{fig:GenAm1B1b3Fit} shows that this approximation is very good even for not too large values of $Z$. 
Evaluating the term $(f_R F_X)^{\frac{1}{2}}$, using the approximation (\ref{eq:XZFit1}), we find that 
\begin{equation}
(f_R F_X)^{\frac{1}{2}}\approx \frac{(2\lambda b)^{\frac{1}{2}}}{(\kappa^2 Z)^{1/3}} \ ,
\end{equation}
which inserted in Eq.~(\ref{eq:r2}) leads to
\begin{equation}\label{eq:r2approx1}
r^2\approx \frac{(\kappa^2 Z)^{1/3}}{(2\lambda b)^{\frac{1}{2}}W^2}\approx \frac{(2\mu\eta)^{2/3}}{(2\lambda b)^{\frac{1}{2}}} e^{\frac{(4M-2\eta)y}{3}}\ .
\end{equation}
From this expression it follows that for astrophysical objects, which should have $M^2\gg \mu^2$, such that $\eta\approx M$,   the area function $r^2(y)$ grows to infinity as $y\to \infty$. This is in clear contrast with what we observe in the GR case, where $r^2\approx (2\eta)^2e^{-2(\eta-M)y}$ tends to zero in that limit. We thus see that $r^2$ grows as we move away from the central object ($y\to 0$ limit) but also grows as the center is approached ($y\to \infty$ limit). This property of the two-spheres indicates that such solutions have the structure of a wormhole~\cite{Visser:1995,Bronnikov:2003,Lobo:2008,Lobo:2017,Delhom:2019}. This is verified in Fig.~\ref{fig:R2WH}, where the function $r^2$ is plotted in three different cases, using the exact analytical expressions and is compared with the approximation (\ref{eq:r2approx1}). 

With the approximation (\ref{eq:XZFit1}), we can evaluate the line element in the limit $y\to\infty$, obtaining
\begin{eqnarray}
ds^2&=&-\frac{\mu^{2/3}e^{\frac{4(\eta-2M)y}{3}}}{(2\lambda b)^{1/2}(2\eta)^{4/3}}dt^2+r^2(y)d\sigma^2\nonumber \\ &+& \left(\frac{(2\eta)^{8/3}e^{\frac{4(M-2\eta)y}{3}}}{(2\lambda b)^{1/2}}-2a\kappa^2\right)dy^2 \ ,
\end{eqnarray}
where the exponential term in front of $dy^2$ should be neglected if $a\neq 0$, thus leading to a simpler expression. Note also that this solution only makes sense when $a<0$, since otherwise the metric can change its signature. Rewriting this line element in terms of the variable $r$ instead of $y$, we end up with 
\begin{equation}\label{eq:WHr}
ds^2=-\frac{\mu^2}{(2\lambda b)^{3/2}r^4}dt^2+\frac{18|a|\mu^2}{(2M-\eta)^2r^2}dr^2+r^2d\sigma^2 \ .
\end{equation}
It is worth noting that this line element has the same functional dependence (up to constants) as the one found in Ref.~\cite{Afonso:2017aci} by directly coupling a free scalar field to the EiBI theory of gravity and solving the corresponding field equations. In that case one also obtains an (asymmetric) wormhole configuration with this type of (approximated) internal line element. In contrast, if one maps the JNW solution (\ref{eq:wyman}) from GR into the EiBi theory, one does not find wormhole solutions of the type we have found here \cite{Afonso:2019fzv}. Thus, the quadratic gravity model coupled to a quadratic scalar field (studied here by mapping the JNW solution) is phenomenologically closer to a free scalar field minimally coupled to EiBI (via direct resolution of the equations) than to the solution found via mapping of JNW into EiBI. 

Using the approximated expression (\ref{eq:WHr}) to evaluate curvature invariants in the $r\to\infty$ limit (on the other side of the wormhole), we find that \begin{eqnarray}
R(g)&=& -\frac{6}{\rho}+\frac{2}{r^2} \, , \\
R_{\mu\nu}(g)R^{\mu\nu}(g)&=& \frac{36}{\rho^2}+\frac{2}{r^4} \, , \nonumber\\
{R^\alpha}_{\mu\nu\beta}{R_\alpha}^{\mu\nu\beta}&=& 4\left(\frac{27}{\rho^2}+\frac{1}{r^4}-\frac{2}{\rho r^2}\right)\nonumber \ ,
\end{eqnarray}
which are all finite as $r\to \infty$ (here we defined $\rho\equiv \frac{18|a|\kappa^2}{(2M-\eta)^2}$). \\
The geodesic equation for a line element of the form $ds^2=-C(x)dt^2+D(x)^{-1}dx^2+r^2(x)d\sigma^2$ can be written as 
\begin{equation}\label{eq:geodesics_gen}
\frac{C(x)}{D(x)}\left(\frac{dx}{d\tau}\right)^2=E^2-C(x)\left(\frac{L^2}{r^2(x)}-k\right) \ ,
\end{equation}
where $E$ and $L$ are constants of the motion, $\tau$ is an affine parameter, and $k=1,0,-1$ for space-like, null, and time-like geodesics, respectively. Inserting the line element (\ref{eq:WHr}) in this equation, with $x=r$, we find that for large $r$ 
\begin{equation}\label{eq:geodesics_WH}
\left(\frac{dr}{d\tau}\right)^2=\frac{E^2(2M-\eta)^2(2\lambda b)^{3/2}}{18|a|\mu^4}r^6+O(r^2) \ ,
\end{equation}
which indicates that all geodesics rapidly degenerate into radial null geodesics, regardless of their $L^2$ or $k$. Integration of Eq.~\eqref{eq:geodesics_WH} leads to 
\begin{equation}
r^2=\pm \frac{3\sqrt{2|a|}\mu^2}{(2M-\eta)(2\lambda b)^{3/4}}\frac{1}{2E(\tau_0-\tau)} \ ,
\end{equation}
where the $\pm$ sign denotes outgoing and ingoing geodesics. From this result it is evident that any outgoing geodesic will reach infinity at a finite affine parameter $\tau_0$, while incoming geodesics must have originated at a finite affine instant. As a consequence, this geometry must be regarded as singular, despite having its curvature invariants bounded everywhere.\\

\begin{figure}
\centering
\includegraphics[width=0.45\textwidth]{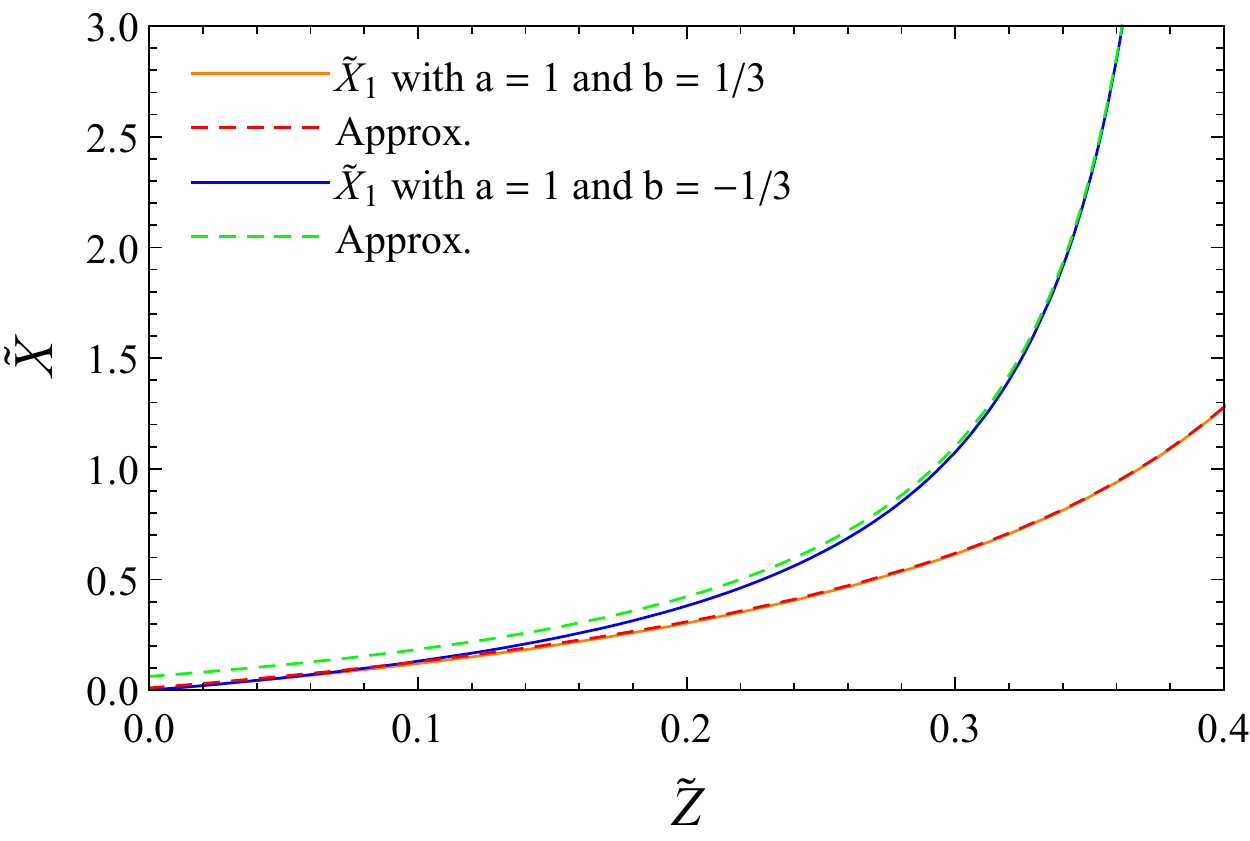}
\caption{Comparison of the type of solution $\tilde{X}(\tilde{Z})$ presented in Fig.~\ref{fig:xz2}, with the analytical approximation (\ref{eq:XZFit2}).}
\label{fig:GenA1B1b3Fit}
\end{figure}

\begin{figure}
\centering
\includegraphics[width=0.45\textwidth]{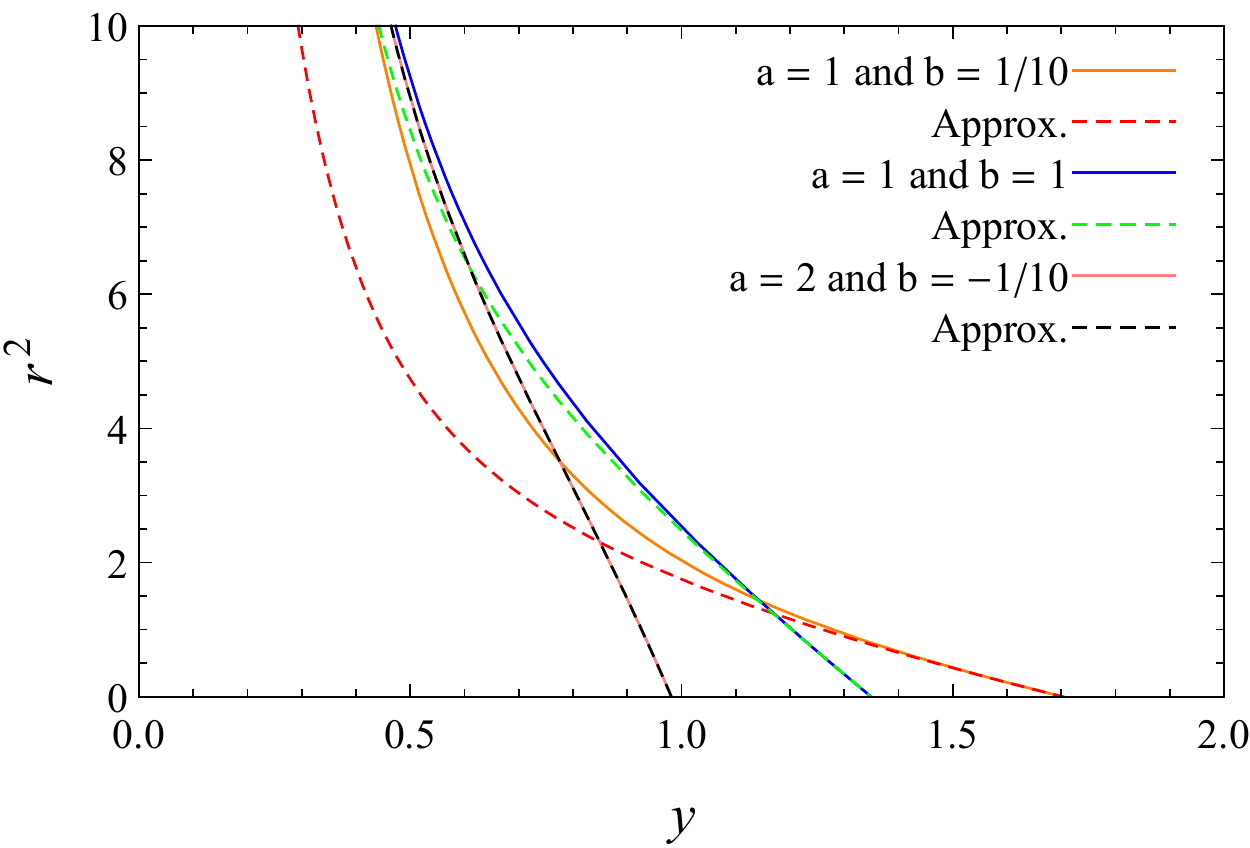}
\caption{The radial function $r^2(y)$ reaches zero at a finite value of $y$. The analytical approximation near the maximum value of $y$ is also shown (dashed curves). As $y\to 0$  the (solid) curves converge to the GR solution, but are not shown here for a better illustration of the end points. }
\label{fig:R2CompactBall}
\end{figure}

Analytical approximations can also be obtained in the cases shown in Fig.~\ref{fig:xz2}, for which one finds
\begin{equation}\label{eq:XZFit2}
\tilde{X}(\tilde{Z})\approx -\frac{8a(a+3b)+15 b^2}{8a(a+b)(2a+3b)}-\frac{(2a+3b)}{16a^4 \tilde{Z}_c(\tilde{Z}-\tilde{Z}_c)} \ ,
\end{equation}
where $\tilde{Z}_c={\sqrt{a+b}}/{2 a^{3/2}}$. One can see that, as shown in Fig.~\ref{fig:GenA1B1b3Fit}, this expression is  an excellent approximation. The behavior of $r^2(y)$ as $y\to \infty$ is then dominated by 
\begin{equation}\label{eq:r2approx2}
r^2\approx \frac{4a^2(\tilde{Z}_c-\tilde{Z})}{(2a+3b)W^2}\ ,
\end{equation}
 which clearly goes to zero as $Z\to Z_c$ at a finite value of $y$. The representation of $r^2(y)$, together with its corresponding analytical approximation near the end point, is presented in Fig.~\ref{fig:R2CompactBall}. The line element in the region near $\tilde{Z}_c$ takes the approximated form
 \begin{equation}
ds^2=\frac{4a^2(\tilde{Z}_c-\tilde{Z})}{(2a+3b)}\left(ds^2_{GR}+\frac{2\mu^2 b\sqrt{a}}{\sqrt{a+b}}dy^2\right) \ ,
\end{equation}  
 which is clearly pathological, because it collapses to zero at some finite $y=y_c$, when $\tilde{Z}_c-\tilde{Z}=0$. Though further analytical discussions around this region are difficult (because we might be dealing with small values of $y$ that do not allow to consider certain simplifications of the expressions involved), a glance at  Fig.~\ref{fig:R2CompactBall} suggests that $r^2(y) \propto (\tilde{Z}_c-\tilde{Z}) \propto (y_c-y)$, i.e., it goes to zero linearly. This allows one to get an approximation for the geodesic equation, which will be dominated by a term of the form 
 \begin{equation}
\left(\frac{dy}{d\tau}\right)^2\approx \frac{\tilde{\alpha}^2}{(y_c-y)^2} \ ,
 \end{equation} 
 where $\tilde{\alpha}$ is some positive constant. 
 Integrating this equation, it follows that $y=y_c$ is reached in a finite affine time, which confirms the geodesic incompleteness of this solution. 
 
\section{Summary and conclusion}\label{sec:summ}
We have extended the correspondence that exists between the space of solutions of GR and RBGs in the case in which the matter source in the GR frame is a free massless scalar field and the target RBG theory is an arbitrary function of the form $f(R,Q)$, with $R=g^{\mu\nu}R_{(\mu\nu)}$ and $Q=g^{\mu\alpha}g^{\nu\beta}R_{(\mu\nu)}R_{(\alpha\beta)}$. We have shown that the correspondence implies that the form of the scalar matter Lagrangian coupled to the $f(R,Q)$ theory can always be found and  takes the simple form (\ref{eq:matter_lagrangian_P_map}), where $R=\kappa^2 X$ and $Q=R^2$. As a result, starting with a free scalar field solution in GR one can construct new solutions in any $f(R,Q)$ theory by just solving the relation between the kinetic terms $Z=h^{\mu\nu}\partial_\mu\phi\partial_\nu\phi $ and $X=g^{\mu\nu}\partial_\mu\phi\partial_\nu\phi $, provided by Eq.~(\ref{eq:ZofX}) and the definition of $|{\hat\Omega}|$ in Eq.~(\ref{eq:detOm}) and $F_X$ in Eq.~(\ref{eq:FX}), namely,
\begin{equation}
Z=\frac{X}{f_R}\sqrt{1+\frac{2\kappa^2X f_Q}{f_R}} \ .
\end{equation}
As an illustration, we have studied static, spherically symmetric configurations in the quadratic RBG theory (\ref{eq:f_model}), using as seed the JNW solution of GR. Among the possible solutions, we have identified two main cases, one which describes naked singularities and another in which an asymmetric wormhole structure arises. The latter is particularly relevant because it closely captures the geometry already observed in the coupling of a free scalar field to the EiBI gravity theory~\cite{Afonso:2017aci}. 
~In contrast, it is qualitatively different from the solutions found by mapping the JNW solution into EiBI, which does not exhibit wormhole solutions \cite{Afonso:2019fzv}, but other kinds of compact objects.  This puts forward that by mapping known solutions of GR into other gravity theories one can generate configurations that exhibit nontrivial new phenomenology.  From a pragmatic point of view, this indicates that it may be possible to generate a catalogue of configurations which capture the essence of the effective phenomenology of compact objects in RBGs. 

Obviously, in order to make further progress in that direction one should be able to handle more general matter theories than the ones considered here and, in particular, those with a nonzero potential or with nonlinear kinetic terms in the Einstein frame. Such an extension would allow us to map specific problems of RBGs into GR plus modified matter, in much the same way as done in  the study of boson stars in Ref.~\cite{Maso-Ferrando:2021ngp} or in the analysis of rotating electrovacuum solutions in Refs.~\cite{Afonso:2021pga,Shao:2020weq}, thus allowing to address complex problems of physical interest in other RBGs beyond the $f(R)$ family.  The consideration of other matter sources such as fluids and electromagnetic fields is another important issue which we hope to address in the future. 

Turning back to the specific spherical scalar field configurations studied here,  the fact that a wormhole spacetime with bounded curvature scalars can nonetheless be geodesically incomplete, raises questions about the physical mechanisms that make such spacetime singular. A detailed analysis of the tidal forces in this scenario is currently underway and we hope to report on it soon. 

\begin{acknowledgments}
The authors would like to acknowledge 
Funda\c{c}\~ao Amaz\^onia de Amparo a Estudos e Pesquisas (FAPESPA), 
Conselho Nacional de Desenvolvimento Cient\'ifico e Tecnol\'ogico (CNPq)
and Coordena\c{c}\~ao de Aperfei\c{c}oamento de Pessoal de N\'ivel Superior (CAPES) -- Finance Code 001, from Brazil, for partial financial support. This research has also been funded by the Spanish Grant FIS2017-84440-C2-1-P funded by MCIN/AEI/10.13039/501100011033 ``ERDF A way of making Europe'', Grant PID2020-116567GB-C21 funded by MCIN/AEI/10.13039/501100011033, the project PROMETEO/2020/079 (Generalitat Valenciana), and by  the European Union's Horizon 2020 research and innovation programme under the H2020-MSCA-RISE-2017 Grant No. FunFiCO-777740. 
\end{acknowledgments}

{}
\end{document}